# Trace Element Partitioning between CAI-Type Melts and Grossite, Melilite, Hibonite, and Olivine

Gokce Ustunisik[a,b*], Denton S. Ebel[b,c], David Walker[c,b], Roger L. Nielsen[d,a], and Marina Gemma[c,b]

[a]Department of Geology and Geological Engineering, South Dakota School of Mines and Technology, Rapid City, SD, 57701
[b]Department of Earth and Planetary Sciences, American Museum of Natural History, New York, NY, 10024-5192
[c]Department of Earth and Environmental Sciences, Lamont Doherty Earth Observatory of Columbia University, Palisades, NY, 10964-8000
[d]104 CEOAS Admin, College of Earth Ocean and Atmospheric Sciences, Oregon State University, Corvallis, OR 97331

[*]**Corresponding author**: Gokce Ustunisik
**E-mail:** Gokce.Ustunisik@sdsmt.edu

**ABSTRACT**

We determined the mineral-melt partition coefficients ($D_i$'s) and the compositional and/or temperature dependency between grossite, melilite, hibonite, olivine and Ca-, Al-inclusion (CAI)-type liquids for a number of light (LE), high field strength (HFSE), large ion lithophile (LILE), and rare earth (REE) elements including Li, Be, B, Sr, Zr, Nb, Ba, La, Ce, Eu, Dy, Ho, Yb, Hf, Ta, Th. A series of isothermal crystallization experiments was conducted at 5 kbar pressure and IW+1 in graphite capsules. The starting compositions were selected based on the calculated and experimentally confirmed phase relations during condensation in CI dust-enriched systems (Ebel and Grossman, 2000; Ebel, 2006; Ustunisik et al., 2014).

The partition coefficients between melt and gehlenite, hibonite, and grossite show that the trace element budget of igneous CAIs is controlled by these three major Al-bearing phases in addition to pyroxene. In general, LE, LILE, REE, and HFSE partition coefficients (by mass) decrease in the order of $^{\text{Gehlenite-Melt}}D_i > {}^{\text{Hibonite-Melt}}D_i > {}^{\text{Grossite-Melt}}D_i$. The partition coefficients between gehlenitic melilite and CAI melt are approximately Be:0.14, B:0.07, Sr:0.79, Zr:0.02, Nb:0.01-0.02, Ba:0.05, La:1.03-3.44, Ce:1.2-3.86, Eu:1.19-2.88, Dy:1.14-3.23, Ho:1.04-2.91, Yb: 0.7-1.70, Hf:0.02, Ta:0.01-0.02, Th:0.31-1.71. These results suggest that $^{\text{Gehlenite-Melt}}D_i$ vary by a factor of 2-3 in different melt compositions at the same T (~1500 °C). $^{\text{Hibonite-Melt}}D_i$ exhibit a range as Be:0.02-0.04, B:0.01, Sr:0.21-0.66, Zr:0.02-0.18, Nb:0.03-0.05, Ba:0.02-0.06, La:0.56-4.38, Ce:0.52-3.54, Eu: 0.33-0.84, Dy: 0.25-0.32, Ho:0.17-0.29, Yb0.05-:0.19, Hf:0.05-0.38, Ta:0.02, Th:0.31-1.71. Increased melt Al and Ca, relative to earlier work, increases the compatibility of $^{\text{Gehlenite-Melt}}D_i$, and also the compatibility of $^{\text{Hibonite-Melt}}D_i$, especially for La and Ce. $^{\text{Grossite-Melt}}D_i$ of individual mineral-melt pairs are Be: 0.43, Sr: 0.31, Zr: 0.09, Nb: 0.01, Ba: 0.03, La: 0.06, Ce: 0.07, Eu: 0.13, Dy: 0.04, Ho: 0.04, Yb: 0.03, Hf: 0.01, Ta: 0.01, Th: 0.01 for #18 at 1550 °C and

as Sr: 0.29, Nb: 0.03, La: 0.07, Ce: 0.09, Eu: 0.10, Dy: 0.05, Ho: 0.04, Yb: 0.02, Hf: 0.003, Ta: 0.02, Th: 0.02 for #7 at 1490 °C.

Olivine partitioning experiments confirm that olivine contribution to the trace element budget of CAIs is small due to the low $^{\text{Olivine-Melt}}D_i$ at a range of temperatures while $^{\text{Olivine-Melt}}D_{\text{Eu, Yb}}$ are sensitive to changes in T and $f\text{O}_2$. The development of a predictive model for partitioning in CAI-type systems would require more experimental data and the use of analytical instruments capable of obtaining single phase analyses for crystals < 5µm.

*Keywords*: condensation; trace element partitioning; CAI; melilite; grossite

## 1. INTRODUCTION

The abundance and distribution of trace elements in calcium- and aluminum- rich inclusions (CAIs) from chondritic meteorites, especially the igneous CAIs, can provide valuable information on high temperature processes in the primitive solar nebula. These processes include volatilization and fractional condensation, changes in the temperature and oxygen fugacity ($f\text{O}_2$) of nebular gas reservoirs, and exchange of trace elements between inclusions and vapor. Interpretation of trace element patterns in CAIs is hampered due to our poor understanding of the temperature, pressure, and compositional dependencies of the partition coefficients, particularly in FeO-poor systems. For some phases (e.g., grossite, $CaAl_4O_7$) there is no experimental data whatsoever. A systematic assessment of the partitioning behavior of a wide range of trace elements and REEs between Ca-, Al-rich melts and their stable phase assemblages at a range of temperature and bulk compositions is needed.

Interpretation of the trace-element evolution of any system requires that we understand how each trace element of interest behaves. This is complicated by the fact that trace element partitioning behavior is dependent on a number of parameters, including pressure, temperature, $f\text{O}_2$ and melt composition. On Earth, differences in trace element concentrations among samples from a given volcanic center can be assessed and attributed to processes such as fractional crystallization, partial melting, etc. by measuring the trends exhibited in the composition of the evolving magmatic system. The context of lava stratigraphy often is an aid in understanding the evolutionary process, at least temporally. A quantitative understanding of how those processes drive changes in trace element composition allows us to develop more precise models for how magmatic systems operate. This analysis usually entails measurement of phase composition, petrographic information, and an understanding of the phase equilibria. For terrestrial systems, in disentangling information on the character of the array of progressive magmas (differentiated liquids from the same starting composition), partition coefficients for all fractionating phases and their phase equilibria often exist.

For extraterrestrial systems, interpretation of trace element systematics can be more complex. For example, modeling variations in trace and rare earth element (REE) concentrations within individual CAIs is made more difficult because the progressively evolved liquids that we often see in terrestrial systems are absent. This is due to CAI's cumulate or granoblastic textures, with a potentially multistage petrogenesis, and the complexities that affect open system processes (Beckett et al., 2006). Such processes include 1) disequilibrium, rapid crystal growth (Kennedy et al., 1997), 2) sector zoning (Hutcheon et al., 1978; Steele et al., 1997), 3) progressive crystallization from an evolving liquid effected by volatilization (e.g., Richter et al., 2002) or incomplete dissolution (Kennedy et al., 1997), 4) subsolidus equilibration with other primary phases, 5) diffusive homogenization (Meeker et al., 1983; Connolly et al., 2003), 6) multiple heating and secondary alteration events (MacPherson and Davis, 1993; Beckett et al., 2000), and 7) processes leading to the entrapment of small inclusions of trace-element-rich phases in the analyzed volume of what one thinks is a homogeneously crystallizing major phase like melilite. (Paque et al., 2016). In addition, unlike terrestrial differentiation, the separate aliquots of bulk composition produced between successive intervals of crystallization inside a CAI would not be available for analysis, as they are in the successive products erupted from a terrestrial volcano. In the case of CAIs, we have only the end product. Progressive changes in melt composition can only be inferred by examining trace element zoning profiles within the crystallizing phases. Yet, those phases may or may not have been in equilibrium with the nearby melt. This obviously obstructs the application of experimentally determined partitioning data to understanding the abundances in natural CAIs. One way to get around this problem is to measure the partition coefficients of elements of interest between CAI-type melts and their coexisting phases under controlled experimental conditions where we can apply systematic compositional and temperature constraints from which equilibrium melt compositions can be calculated and compared to typical CAI-type melts which are associated with the refractory oxides (hibonite, grossite) and silicates (olivine, melilite) to develop quantitative insight into the processes that may shift them from chemical equilibrium.

We conducted a series of isothermal crystallization experiments using a number of bulk compositions from Ustunisik et al. (2014) to determine partitioning of trace and REEs in specific fields of liquid-stable condensation space (Ebel, 2006) between grossite, melilite, hibonite, olivine, and CAI-type liquids. Hibonite, grossite, melilite and olivine are stable in specific, distinct ranges of natural composition, as determined by phase equilibria. Therefore, the ranges of composition of equilibrium liquids are limited, complicating our ability to calibrate expressions describing compositional and temperature dependence of partitioning in CAI-type melts. Nevertheless, our goal in this investigation was to expand the range of existing temperature and composition compared with the previously used starting liquids such as CAI-B of Beckett et al. (1990) on which most of the previous trace and REE partitioning experiments were conducted. From our new equilibrium partition coefficients, we can evaluate the degree to which the phases in natural CAIs are in equilibrium with each other (by calculating the mineral/mineral equilibria from the mineral/melt partition coefficients and comparing them for the different minerals).

## 2. PREVIOUS WORK

For a complete understanding of the petrogenesis of CAIs we need to isolate specific parts of the system where the behavior of a set of trace elements may be more predictable. Previous experimental studies by Stolper (1982) and Richter et al. (2002) showed that Type B (igneous) CAIs in carbonaceous chondrites may represent closed systems with respect to some refractory elements (e.g., Ca, Al, Ti). Therefore, they offer a unique opportunity to understand the distribution of trace elements in controlled conditions. In their experiments on Type B CAIs, MacPherson et al. (1984) and Stolper and Paque (1986) found that such igneous CAI were primarily affected by fractional crystallization from a single starting liquid of a known bulk composition, crystallization sequence, and cooling rate. Their idea initiated a number of phase equilibria experiments on CAIs and their refractory oxides (spinel, hibonite, and perovskite) and silicates (melilite, Al-, Ti-bearing clinopyroxene, olivine, and plagioclase) in order to constrain the roles of volatilization and fractional condensation, the temperature and oxygen fugacities of nebular gas reservoirs, and the trace element budget of the primitive solar nebula. The majority of these phase equilibria experiments focused on a small number (compared to that normally analyzed in natural samples) of trace and rare earth element components and were limited to phases such as hibonite (Drake and Boynton, 1988; Kennedy et al., 1997); melilite (Beckett et al., 1990); synthetic olivine in CMAS system (Evans et al., 2008) and more importantly to a single bulk liquid composition (e.g., CAI-B of Beckett et al., 1990). Significantly, there have been no partitioning experiments performed for grossite, a rare but important phase in CAIs (MacPherson, 2004; Beckett et al., 2006).

Kennedy et al. (1997) focused on understanding the genesis and evolution of hibonite in compact Type A (CTA) CAIs by comparing the equilibrium partitioning behavior of Sr, La, Ce, Eu and Hf at 1400°C and 1350°C low pressure and IW-1.5. They concluded that igneous hibonite should not have equilibrium $^{Min-Melt}D_{LREE} > {}^{Min-Melt}D_{HREE}$, where $^{Min-Melt}D_i$ are partition coefficients by mass $M_i$(mineral)/$M_i$(melt) (Beattie et al., 1993), implying that the LREE-enriched hibonites in some CTA CAIs from Leoville and Allende (both CV3 chondrites) are inconsistent with hibonite being generated from the melts that formed these inclusions and therefore they must be relicts that survived the most recent partial melting event.

In the absence of information required to calculate the activity of trace components, most investigators assume ideality and apply simple mineral/melt partition coefficients. However, existing data indicates that the partitioning behavior of most trace elements is dependent on melt composition. Therefore, reliance on simple mineral/melt partition coefficients leaves our understanding of the details of condensation effects on REE incomplete at best. Drake and Boynton (1988) investigated the partitioning of selected trace and REEs (Li, Sr, La, and Eu) between hibonite and melt as a function of oxygen fugacity (air and NNO) at 1470°C. They assumed that hibonite might be an important host for REEs in carbonaceous chondrites and therefore may control trace element budgets during condensation. In detail, hibonite along with perovskite and melilite were found to have a strong preference for LREEs relative to HREEs. They estimated what they referred to as the activity coefficient ratios of $Eu^{+3}/Eu^{+2}$ in hibonite and melt at 1470°C by assuming $Eu^{+2}$ behaved identically to Sr and inferred the behavior of $Eu^{+3}$ by interpolating between the behavior $Gd^{+3}$ and $Sm^{+3}$. They then, by inference, calculated the activity coefficient ratios of other REEs in hibonite and melt at the same temperature. Their results showed a significant departure from ideality in the mixing properties for Eu.

Kennedy et al. (1994, 1997) found similar $^{Hibonite-Melt}D_i$ values for Sr and Eu at the experimental conditions of Drake and Boynton (1988). In the case of $^{Hibonite-Melt}D_{La}$ the experimental results differ between the two studies, but these differences cannot be assessed due to several factors. These include the fact that each study used different bulk compositions, including differences in refractory oxides (CaO, $Al_2O_3$ and $TiO_2$), variations in temperature and $fO_2$, and significant differences in doping levels (wt. % vs. ppm) due to the use of different analytical procedure (EPMA vs. SIMS) all of which certainly affect $^{Hibonite-Melt}D_{REE}$ (Fig. 1).

Melilite has received the most attention in trace element partitioning studies in CAIs. Nagasawa et al. (1980) showed that the $^{Melilite-Melt}D_{REE, Sr, Sc}$ for the temperature range 1500-1542°C under atmospheric and low $fO_2$ conditions (IW+1) depends on composition with behaviors ranging from strongly compatible to incompatible. They found a modest decrease in partition coefficients from the LREE to the HREE. They attributed this relatively flat pattern in the REE partitioning behavior to the distortion of the Ca site in the melilite crystal structure. Specifically, the distortion of the octahedral site wherein four of the Ca-O bonds are much weaker or longer than the other four resulting in a "loose" coordination around the Ca site. Therefore, partitioning of REE into the melilite lattice appears to be only weakly dependent on the ionic radius of the specific REE.

Beckett et al. (1990) measured the partitioning behavior for Be, Sc, Ba, La, Ce, and Tm between melilite and melt during the crystallization of analogs to Type B1 inclusions (igneous, with zoned melilite-rich mantles) from CV3 carbonaceous chondrites over the temperature range 1316-1193°C at IW. Their results showed that the $^{\text{Melilite-Melt}}D_{\text{Be, Sc, Ba, La, Ce, Tm}}$ within a CV3 Type B1 CAI could not solely reflect equilibrium partitioning. This was attributed to the compositional dependence of the partitioning of these elements between Åk$_{25-65}$ (zoned) melilite and melt, the $^{\text{Melilite-Melt}}D$ of divalent cations (substituting into the melilite M1 site) being approximately constant. However, trivalent cations (substituting into the tetrahedral T1 site) have partition coefficients that are negatively correlated with increasing mole fraction Åkermanite ($X_{\text{Ak}}$) or decreasing temperature. In addition, Beckett et al. (1990) demonstrated that Be becomes compatible with increasing $X_{\text{Ak}}$ or decreasing temperature while Ba and Sc substitute on the Ca site but are incompatible and essentially independent of $X_{\text{Ak}}$ in the CAI-like bulk compositions of their experiments.

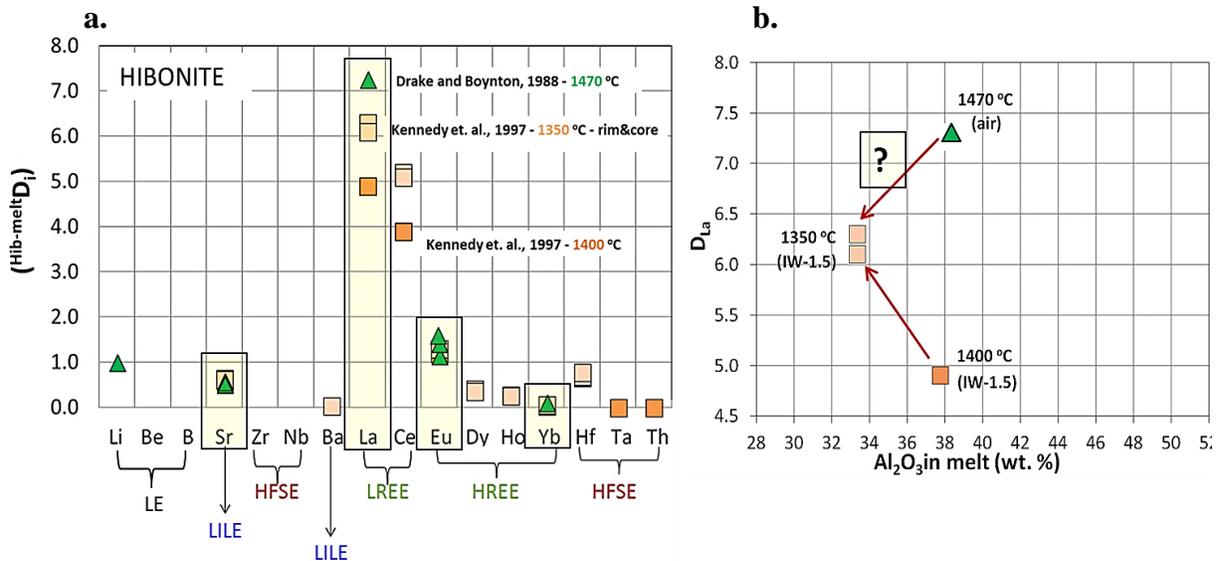

**Figure 1. a.** Hibonite-Melt partition coefficient ($^{\text{Hibonite-Melt}}D_i$) for Li, Sr, La, Eu, and Yb from Drake and Boynton (1988, green filled triangles at 1470 °C, air) and for Sr, Ba, La, Ce, Eu, Dy, Ho, Yb, Hf, Ta, and Th from Kennedy et al. (1997). orange filled squares, light and dark colors are at 1350 °C, 1400 °C and IW-1.5 respectively). **b.** $^{\text{Hibonite-Melt}}D_{\text{La}}$ versus Al$_2$O$_3$ content (wt%) in melts from Drake and Boynton (1988) and Kennedy et al. (1997). The effects of temperature, composition and/or $f$O$_2$ on $^{\text{Hibonite-Melt}}D_{\text{La}}$ are difficult to assess independently. **LE**: Light Elements, **LILE**: Large Ion Lithophile Elements, **HFSE**: High Field Strength Elements, **LREE**: Light Rare Earth Elements, **HREE**: Heavy Rare Earth Elements.

Kuehner et. al. (1989) studied $^{\text{Melilite-Melt}}D_{\text{REE, Sr, Y, Zr}}$ between two end-member compositions, akermanite and gehlenite, at 1474°C and 1374°C and IW-2 conditions. They concluded that the general relationships described by Beckett et al. (1990) between REE$^{+3}$ and akermanite content may well be correct for melilite of intermediate compositions but do not hold true for very low $X_{\text{Ak}}$ (< 0.1). They also found that $^{\text{Melilite-Melt}}D_{\text{Sr}}$ is strongly correlated with $X_{\text{Ak}}$. Davis et. al. (1996) investigated how $^{\text{Melilite-Melt}}D_{\text{REE}}$ varied as a function of cooling rate. They found that a fast cooling rate (2°C/h) cannot explain observed REE patterns in Allende CAIs. They showed that $^{\text{Melilite-Melt}}D_{\text{REE}}$ increases as a function of decreasing $f$O$_2$, correlated with the evaporation of Mg and Si from the melt, resulting in the concentration of the refractory trace elements and post-crystallization diffusive re-equilibration altering the trace element abundances without changing the melilite composition. Lundstrom et. al. (2006) concentrated on the distribution of Li, Be and B in addition to a larger set of trace elements and REEs between melilite and the Beckett et al. (1990) CAI-B melt composition. Using their experimentally determined light element partition coefficients and the Li, Be, B systematics, they inferred that fractional crystallization was the primary process, but only within the unperturbed portions of CAIs from Allende. Their $^{Melilite-Melt}D_{Be}$ agreed with Beckett et al. (1990), Be being the only common light element investigated between the two studies. Meteoritic melilite is a $Al_2[MgSi]_{-1}$ binary solid solution requiring charge compensating substitution for any aliovalent cations. Therefore, one might expect a dependence of $^{Melilite-Melt}D_i$ on the MgO, $SiO_2$, and $Al_2O_3$ contents of equilibrium liquids, thus also on the $X_{Ak}$ of melilite at fixed temperature (T).

Developing systematic constraints on $^{Melilite-Melt}D_i$ using existing experiments is difficult since the partitioning data for the commonly studied elements is limited to Be, Sr, Ba, La, Ce, Eu, and Yb. A full set of elements including Li and B are only available from Lundstrom et. al. (2006); yet for their study, temperatures do not exceed 1474 °C, melilite is limited to the compositional range $Ak_{30-70}$, and starting compositions and $fO_2$ are exactly the same as in Beckett et al. (1990) (Fig. 2). Moreover, the effect of gehlenitic melilite ($Ak_{0-10}$) on $^{Melilite-Melt}D_{REEs}$, especially La, is unknown for both Beckett et al. (1990) and other starting compositions as indicated earlier by Kuehner et. al (1989).

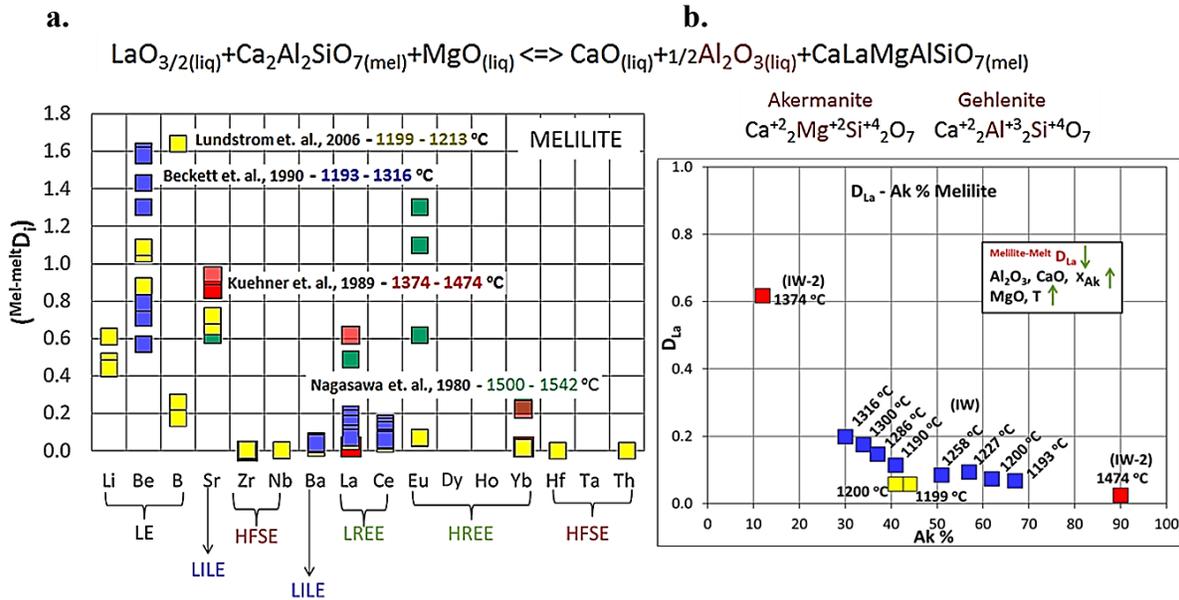

**Figure 2. a.** Melilite-Melt partition coefficient ($^{Melilite-Melt}D_i$) for Li, Be, B, Sr, Zr, Nb, Ba, La, Ce, Eu, Yb, Hf, and Th from Lundstrom et al. (2006, yellow filled squares at 1199–1213 °C, IW); for Zr, La, Sr, and Yb from Kuehner et al. (1989, red filled squares at 1374–1474 °C, IW); for Be, Sc, Ba, La, Ce, and Tm from Beckett et al. (1990, blue filled squares 1193–1316 °C, IW); and for Sr, Sc, La, Eu from Nagasawa et al. (1980. green filled squares at 1500–1542 °C, atmospheric and IW + 1). **b.** Effect of akermanite content of melilite (Ak%) on $^{Melilite-Melt}D_{La}$ from all experiments. $^{Melilite-Melt}D_{La}$ increases with decreasing temperature and MgO while it decreases with decreasing $Al_2O_3$, CaO, and akermanite mole fraction ($X_{Ak}$) for various temperature and $fO_2$. **LE**: Light Elements, **LILE**: Large Ion Lithophile Elements, **HFSE**: High Field Strength Elements, **LREE**: Light Rare Earth Elements, **HREE**: Heavy Rare Earth Elements.

Olivine is a major phase in terrestrial magmatic systems and its partitioning behavior has received considerable attention (Nielsen et al., 2015 and references therein). However, with a few exceptions (e.g., Mallman and ONeill's 2013 investigation of $^{Olivine-Melt}D_{Sc}$, $D_V$, and $D_Y$ as a function of $fO_2$, $^{Olivine-Melt}D_{Ca}$ of Libourel 1999, and some transition elements - e.g., Watson, 1976) essentially all existing data is for Fe-bearing systems (Bédard, 2005). Evans et al. (2008) determined the partition coefficients for a range of REEs and Y, Sc, Al and Zr for synthetic forsterite and ten melt compositions in the CaO-MgO-Al$_2$O$_3$-SiO$_2$ (CMAS) system at 1 bar and 1400°C. Their work showed that $^{Olivine-Melt}D_{Y, Ho}$ are inversely correlated with SiO$_2$ in the melt. They attributed this to strong interaction between REE-O$_{1.5}$ and SiO$_2$ in the melt. Similar to the situation for hibonite and melilite, assessing the temperature and compositional effects on partitioning of trace and REEs between olivine and CAI-like melts is hampered by the fact that existing experiments are restricted to FeO-free bulk compositions (high in CaO and Al$_2$O$_3$) at a single temperature (1400°C; Fig. 3).

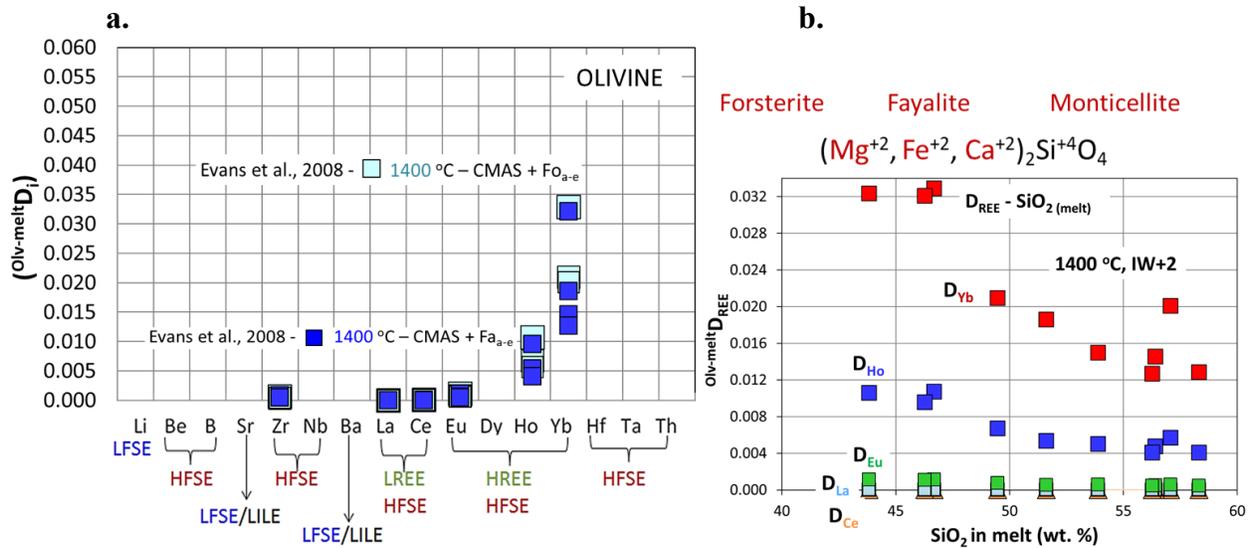

**Figure 3. a.** Olivine-Melt partition coefficient ($^{Olivine-Melt}D_i$) for Zr, La, Ce, Eu, Ho, and Yb from Evans et al. (2008, blue filled squares, light and dark colors are both at 1400 °C at $fO_2$ conditions of air with Fo-bearing at IW + 1 with Fa-bearing olivine respectively). **b.** Variable degrees of decreases in $^{Olivine-Melt}D_{REE}$ with increasing SiO$_2$ content of melt. **LE**: Light Elements, **LILE**: Large Ion Lithophile Elements, **HFSE**: High Field Strength Elements, **LREE**: Light Rare Earth Elements, **HREE**: Heavy Rare Earth Elements.

As shown in detail above, trace element partitioning experiments for Type B (igneous) CAIs in carbonaceous chondrites have often been designed to address a specific question. Therefore, many of the existing crystal/melt distribution coefficients were determined over a limited range of temperature and composition. Those conditions constrain the partition behavior as the system is subjected to volatilization, fractional condensation, remelting, fractional crystallization, and subsolidus reequilibration. Although these processes may be mutually exclusive, an accurate assessment of Type B CAI crystallization requires knowledge of mineral-melt partition coefficients ($^{Mineral-Melt}D_i$) for all coexisting phases under conditions appropriate for the system being modeled. As discussed above, since $^{Mineral-Melt}D_i$ only exist for certain trace elements, phases, temperature, and $fO_2$ conditions, and limited melt composition (mostly CAI-B of Beckett et al., 1990), there continues to be a need for systematic assessment of partitioning of a wide range of trace and REEs between minerals and melts.

## 3. METHODS

### 3.1. Experimental design

Each experimental bulk composition synthesized for this investigation represents the sum of the solid(s) and liquid condensates, minus metal, predicted to be in equilibrium with vapor at a specific T at total pressure ($P^{tot}$) of $10^{-3}$ bar, for a system of solar composition enriched in CI-chondrite dust by a dust-enrichment factor ($d$; Ebel and Grossman, 2000). As shown in Fig. 4a, the predicted (Ebel and Grossman, 2000; Ebel, 2006) and experimentally determined (Ustunisik et al., 2014) bulk condensate composition (solid(s) + liquid) at a specific $d$ factor will change with increase or decrease in temperature, since a portion of the total system is calculated to either evaporate or condense to or from the coexisting vapor. Therefore, starting compositions used in this study are the bulk condensate composition calculated to be in equilibrium with vapor at a specific T, for a specific initial total system composition as described in Ustunisik et al. (2014).

Partitioning experiments were performed at 5 kbar (500 MPa) using the bulk compositions in Table 1 in the phase fields of Figure 4a (Ustunisik et al., 2014) with a trace element dopant. High-pressure promoted the retention of the desired volatile trace element abundances in the melt. Phase assemblages of 5 kbar partitioning experiments were found to be the same as in the one atmosphere phase equilibria experiments from Ustunisik et al. (2014) which used the same starting compositions without dopant. Even though the higher pressure might have minimal effect on partitioning behavior of trace elements while not affecting phase assemblages, low pressure would not favor the retention of volatile trace elements in the starting compositions which is the point of these experiments.

Six bulk compositions #7, 18, 6, 17, 12, and 15 (yellow circles in Fig. 4) were mixtures of oxides, silicates, and carbonates ground in ethanol in an automated agate mortar for >1 hour and dried at 175°C under vacuum for 15 minutes to remove any adsorbed water. In the case of bulk compositions with very high Ca/Si ratio (e.g., #7), carbonate ($CaCO_3$) was used instead of silicate ($CaSiO_3$) as a source of CaO. For the decarbonation process, the homogenized mixture was fired in a platinum crucible in a Deltech furnace overnight in air by heating the mixture to 1400°C via three steps (to 810°C at 25°C/min, to 920°C at 1°C/min, and to 1400°C at 25°C/min). The material was sintered but not melted. Green arrows in Figure 1 show two experiments in the liquid+olivine field (9 and 21) using the bulk compositions 12 and 15 respectively at 1477°C and 1397°C. The reasoning behind this is discussed by Ustunisik et al. (2014), who demonstrated that the bulk condensate compositions are nearly identical at different temperature and $d$ along trend lines parallel to phase boundaries, while bulk condensate compositions diverge more strongly at T far above and below the calculated olivine stability limit (red dot-dot-dash line in Fig. 4). This makes it appropriate to use one starting composition (#12 and #15) to approximate the bulk composition at other T (#9 and #21) that correspond to systems with only slightly different total composition (vapor plus condensate).

The starting compositions used for our experiments (Fig. 4b, highlighted in yellow boxes) are significantly higher in CaO and $Al_2O_3$, to stabilize hibonite and melilite, compared to starting compositions used by previous experiments restricted to the CAI-B of Beckett et al. (1990).

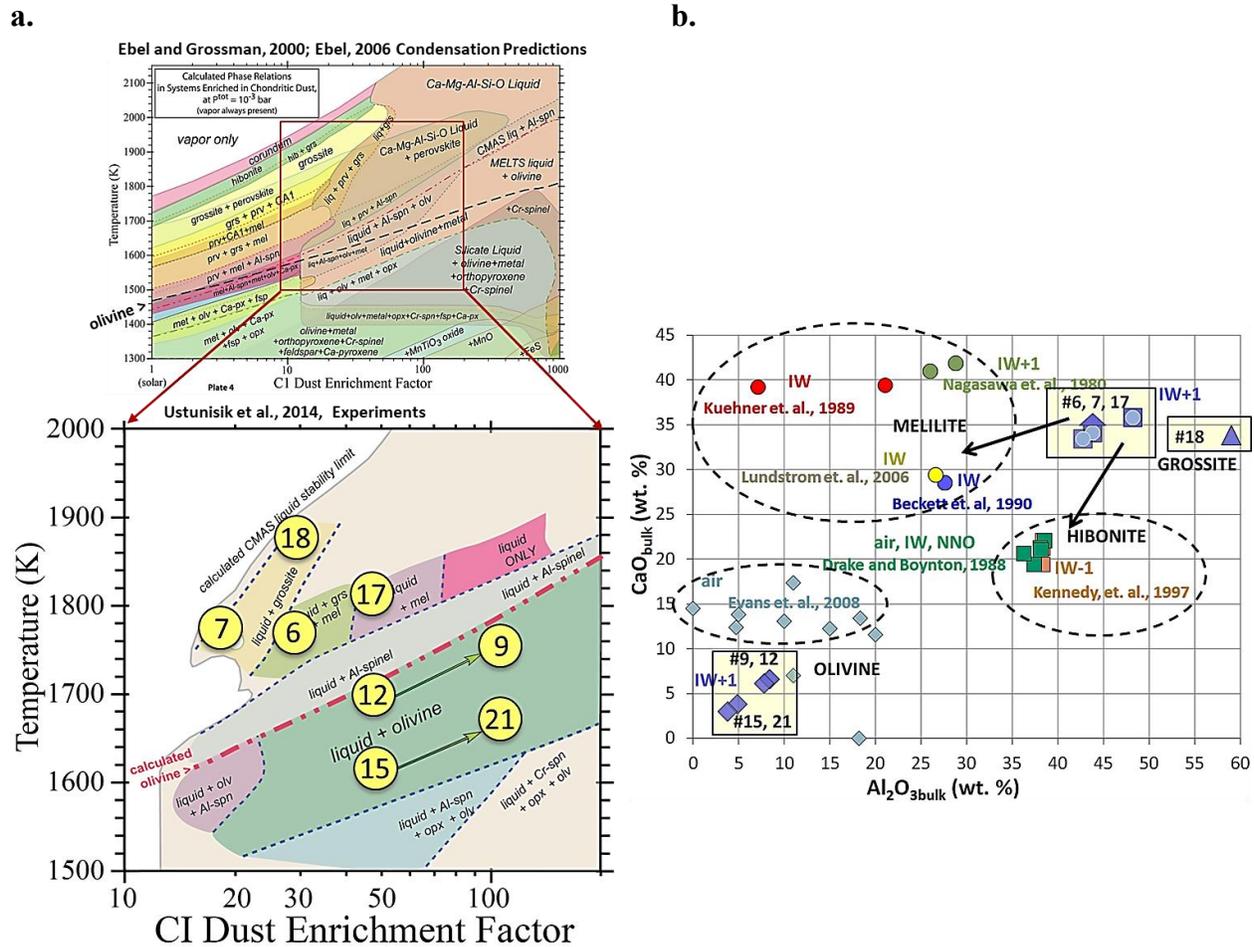

**Figure 4. a.** Area outlined in red square shows experimentally determined phase boundaries (Ustunisik et al., 2014) with respect to calculated phase relations during condensation in CI dust-enriched systems (Ebel and Grossman, 2000; Ebel, 2006). Eight bulk compositions in yellow (#6, #7, #9, #12, #15, #16, #18, and #21) which were used earlier to test predicted liquid + solid assemblages at specific temperatures (T), cooled from a vapor enriched by a specific dust enrichment factor (d, e.g. 20x), in T–d space (Ebel and Grossman, 2000; Ebel, 2006) are used here as starting bulk compositions for trace element, REE partitioning experiments. Along trends indicated by arrows, one bulk composition can be used as a proxy to test other predicted compositions in T-d space. Starting composition #12 is used for composition #9; likewise, #15 is used as the starting composition #21 at the temperature of #21. **b.** Comparison in $Al_2O_3$ vs. CaO space of starting compositions for our experiments with the starting compositions used in previous partitioning experiments ((**b**) dashed circles) for hibonite by Kennedy et al. (1997, at IW-1, orange squares) and Drake and Boynton (1988, in air and at IW and NNO, dark green squares); for melilite by Nagasawa et al. (1980, at IW + 1, light green circles), Kuehner et al. (1989, at IW, red circles), Beckett et al. (1990, at IW, dark blue circle), and Lundstrom et al. (2006, at IW, yellow circle); for olivine by Evans et al. (2008, in air, light blue diamonds). Starting compositions used in our experiments are for hibonite #6, 7, 17 – light blue circles; for melilite #6, 7, 17 – dark blue squares; for olivine #9, 12, 15, 21 – dark blue diamonds; and for grossite #7, 18 – dark blue triangles. The $fO_2$ conditions of our experiments are set to IW + 1.

Our experiments are the first experimental determinations for grossite partitioning and therefore, there are no literature comparisons. In the olivine field, when compared to Evans et al. (2008), our starting compositions are significantly lower in CaO and $Al_2O_3$. To expand the range of conditions for which we have $^{Olivine-Melt}D_i$ data, we used 4 different experiments (#12, 9, 15, 21). Therefore, our experiments not only determine partitioning of trace and REEs between melilite, hibonite, grossite, olivine and CAI-type liquids but also add significant constraints to the compositional and temperature dependence of trace and REE partitioning between olivine and liquid by extension to CaO- and $Al_2O_3$-depleted melts (compared to previous experiments).

The dopant mixture used in these experiments was prepared from oxides of La, Ce, Eu, Dy, Ho, Tb, Yb, Zr, Hf, and Ta; carbonates of Sr and Ba; nitrates of Cs and Rb; Th in the form of thorianite $ThO_2$; Be as the mineral beryl (AMNH collection #40565 from Mogok), and B as boric acid $H_3BO_3$. This dopant mix was added to the starting composition of each experiment in several steps to produce 50-100 ppm final concentrations of each trace element, keeping the total trace + REE concentration <1500 ppm so as not to alter the phase equilibria.

**Table 1.** Bulk compositions and experimental conditions for #6, 17, 7, 18, 9, and 12 along with the stable phases at 5 kbar. Phase abundances are determined optically. Hib: hibonite, Gr: grossite, Geh: gehlenite, Liq: liquid, Olv: olivine.

| Run # | 6 | 17 | 7 | 18 | 9 | 12 |
|---|---|---|---|---|---|---|
| $SiO_2$* | 19.13 | 20.37 | 13.57 | 5.09 | 41.28 | 40.61 |
| $TiO_2$ | 1.86 | 1.75 | 2.05 | 1.89 | 0.34 | 0.37 |
| $Al_2O_3$ | 43.81 | 42.75 | 48.22 | 59.03 | 7.79 | 8.36 |
| FeO | 0.00 | 0.00 | 0.00 | 0.00 | 0.27 | 0.20 |
| MgO | 1.08 | 1.71 | 0.37 | 0.18 | 44.10 | 43.75 |
| CaO | 34.12 | 33.81 | 35.79 | 33.41 | 6.16 | 6.62 |
| Total | 100.00 | 100.39 | 100.00 | 99.60 | 99.94 | 99.91 |
| Duration (hr) | 1 | 1.5 | 1 | 1 | 100 | 111 |
| Temperature (T, °C) | 1500 | 1527 | 1490 | 1590 | 1477 | 1420 |
| Phases | hib + gr | hib | geh+gr | | | |
| | geh+liq | geh+liq | hib+liq | gr+liq | olv+liq | olv+liq |
| Abundance of Phases (%) | hib: 15 | hib: 40 | geh: 45 | gr: 50 | olv: 60 | olv: 70 |
| | gr**: 2 | gr: 25 | gr: 25 | liq: 50 | liq: 40 | liq: 30 |
| | geh: 30 | geh: 10 | hib: 5 | | | |
| | liq: 53 | liq: 50 | liq: 25 | | | |
| *All oxides are reported in wt.%. | | | | | | |
| **Grossite is so fine to analyze, therefore its major or trace element compositions cannot be reported. | | | | | | |

### 3.2. Experimental details

The high P-T experiments were conducted using an end-loaded piston-cylinder apparatus at Lamont Doherty Earth Observatory (LDEO). End loads of 150 tons were applied to the pressure vessel. Piston loading by a 3" ram to reach 5 kbar of 198 bars oil pressure used the calibration for $BaCO_3$ of Fram and Longhi (1992). One experiment was run on each of eight T - bulk composition conditions (Table 1). Run times ranged from 1 to 265 hours. Runs were performed in a ½ inch (″) inside diameter Boyd-England compound pressure vessel with cylindrical Pb-wrapped $BaCO_3$/MgO pressure medium (0.487″, 0.313″, 1.25″), graphite furnace (0.312″, 0.25″, 1.25″), high-density $Al_2O_3$ sleeve (0.248″, 0.181″, 0.25″), one-hole and solid MgO spacers (0.5″), MgO wafer (0.078″), and graphite sample capsule (0.180″ with 0.125" internal cavity), with continuous external $H_2$-$N_2$ (5% $H_2$, 95% $N_2$) flow to protect the thermocouple from oxidation. Experiments were pressurized cold then heated incrementally to 850°C for over an hour after which the assembly was left at 850°C overnight to stabilize and close porosity in the pressure media and graphite capsules. After repressurization, T was raised in 50°C steps of approximately 20 minutes to the run T, repressurized after 20 min, left for the experiment duration (Table 1), and quenched by turning off the electric power. Cooling to less than 400°C took approximately 5 seconds. Run products were mounted in epoxy then sectioned and polished for petrographic and geochemical analysis. A temperature gradient of a few degrees in the piston–cylinder experimental configurations led to phase separation of crystals from liquid by thermal migration (Walker, 2000); the solid crystal phase compacted against the cold side as a result of melt migration to the hotter side (Fig. 5).

### 3.3. Analytical procedures

Backscattered electron (BSE) images (Fig. 6), X-ray intensity maps, and mineral and glass compositions of experimental runs were obtained using the Cameca SX100 electron microprobe (EMP) at the American Museum of Natural History (AMNH). BSE images were used to determine textural characteristics to differentiate quench from equilibrium crystals as well as to observe contrast changes based on mean atomic number in the crystals too fine to analyze. Glass (melt) analyses were obtained more than 100 microns from any crystal to avoid both multiphase analyses as well as to minimize the possibility of sampling liquids influenced by quench growth.

Microprobe analysis conditions for minerals were: 15 kV accelerating voltage, 20 nA beam current, focused electron beam (nominally 1 µm), and peak and background counting times of 30-60 s. The same setup was used for the glass analyses, with the electron beam at 1 µm for both the highly crystalline low temperature experiments and crystal-poor high temperature ones. We analyzed for Si, Ti, Al, Fe, Mg, Ca, and Cr. Standards were synthetic oxides and natural minerals (enstatite for Si, synthetic perovskite for Ca and Ti, synthetic corundum for Al, fayalite for Fe, synthetic $MgAl_2O_4$ spinel for Mg and synthetic magnesiochromite for Cr). Tests on several secondary standards were performed throughout the analytical sessions to verify that calibrations did not drift.

Portions of the experimental runs were mapped for x-ray intensity at 15 kV accelerating voltage, and 20nA beam current using five wavelength dispersive spectrometers (WDS: Al, Fe, Ti, Mg, and Ca), an energy dispersive spectrometer (EDS: Al, Si, Ti, Cr, Fe), and a BSE detector. Instrument parameters were chosen to minimize acquisition time and optimize intensities. X-ray intensity maps were then combined into composite red-green-blue (RGB) three element images to facilitate identification of particular phases that are difficult to probe (e.g., Exp #6, grossite growing around quench melilite crystals).

Trace elements were analyzed at LDEO using laser ablation inductively coupled plasma mass spectrometry (LA-ICP-MS). Glass standards NBS 614, 612, and 610 were used for trace elements at ≤4 ppm, ≤40 ppm, and ≤400 ppm respectively. A 40-50µm beam size was used for standards and minerals and 25µm was used for finer crystals (e.g., hibonite). Dwell times were 50 millisec (ms) for Li and Be, 25ms for Rb and Cs, and 10 ms for the other trace elements.

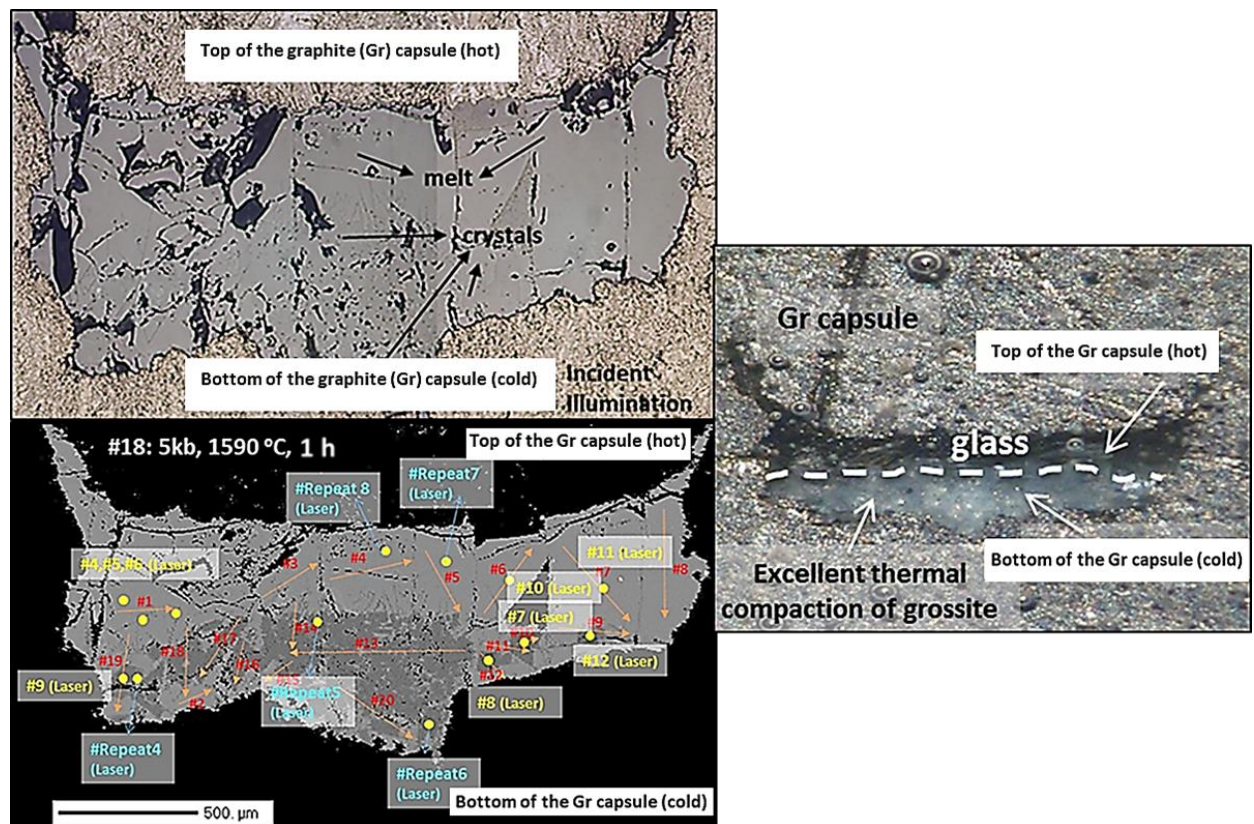

**Figure 5.** Incident light (above) and back-scattered electron (BSE) image (below) along the c-axis of the experiment #18 with the view under binocular microscope (right side). Notations on BSE image indicate EMP (red numbers with orange arrows) and Laser ICP-MS analysis (yellow and blue numbers with yellow dots) locations. Liquid is separating from the thermally compacted grossite crystals at the colder bottom of the graphite capsule.

## 4. RESULTS

We report the results of eight experiments that produced euhedral grossite (Exp. #6, #7, #18), gehlenite (Exp. #6 and #7), hibonite (Exp. #6, #7, #17), and olivine (Exp. #15, #21, #12, and #9) suitable for determining mineral-melt partition coefficients and the bulk composition and/or temperature dependency of $^{\text{Mineral-Melt}}D_i$ (in terms of wt. % or ppm) in temperature vs. dust enrichment factor (T-$d$) space (Fig. 4a).

All runs exhibited some extent of thermal compaction: separation (see Fig. 1 of Yokoyama et al., 2009) of homogeneous glass at the hot top of the capsule and equilibrium solid crystalline phases at the colder bottom (Fig. 5). These experiments grew euhedral and unzoned crystals coexisting with quenched melts (glass) both of which have homogeneous trace element compositions within each charge. No evidence was seen for leakage from the graphite capsules. The concentrations of most trace elements in the melt were in the range of 50-100 ppm. However, Rb and Cs concentrations were below detection limits. We attribute their loss to volatilization during the experiments.

As noted above, analyses of the experiments were performed by EMP and LA-ICP-MS. For major elements, 10-15 analyses were obtained by EMP from each experiment and reported as averages with standard deviations (Tables 2, 3, 4, and 5). This data was used to establish the degree of major element homogeneity of each phase in each experiment. Trace element analyses were performed by LA-ICP-MS on spots where clusters of EMP analyses had been collected (Fig. 6, BSE images). Partition coefficients for individual mineral-melt pairs (not only averages) and average partition coefficients for $^{\text{Grossite-Melt}}D_i$, $^{\text{Gehlenite-Melt}}D_i$, $^{\text{Hibonite-Melt}}D_i$, and $^{\text{Olivine-Melt}}D_i$ were calculated (Tables 2-5) in order to provide a measure of local equilibria for the trace elements. The reported error for the trace elements is based on the counting statistics for individual analyses. This approach is designed to avoid the analytical problems inherent in the application of multiple techniques with different analytical volumes (Nielsen et al., 2017).

In developing our starting compositions, one of our goals was to restrict our experiments to systems with low modal crystallinity (high % liquid) to facilitate equilibration (due to the faster diffusion rates for melt vs. minerals) and our ability to obtain analyses of single phases (i.e., reduce the probability of multiple phases in the analytical volume, cf. Nielsen et al., 2017). In general, that is most easily achieved when only one crystalline phase is present. However, the presence of multiple phases in an experiment allows us to obtain mineral/mineral partitioning data that is useful for cases where no melt is present.

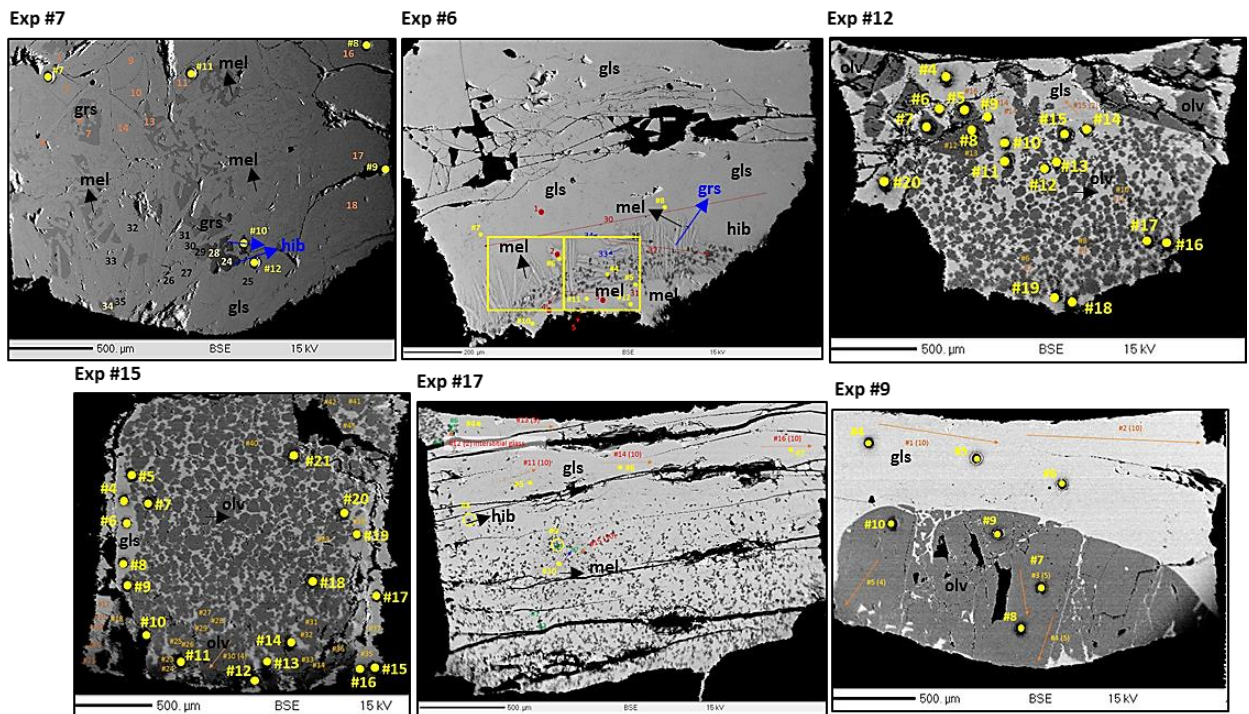

**Figure 6.** BSE image of experiments #6, 7, 9, 12, 15. Labels indicate olv (Mg-olivine), gr (grossite), hib (hibonite), mel (melilite) and gls (glass). Black, red and orange numbers and orange arrows show EMP point and line analyses. Yellow numbers with yellow dots show laser ICP-MS analysis locations. Yellow squares in Exp #6 show the location of X-ray intensity maps which indicates grossite growing around the quench melilite crystals.

## 5. DISCUSSION

In the section below, we will discuss how the different groups of trace elements behave in each of the minerals observed in our experiments. For each phase, we will describe how the light elements (LE), Li, Be, B; the large ion lithophile elements (LILE) Sr, Ba; the REEs, and the HFSE Zr, Hf, Nb, Ta, Th partition within the context of the liquid composition, we will propose site occupancy and substitution mechanisms, and we will compare our data with the published data for the same phases. Changes in the partition coefficients and Onuma diagrams are given in log-scale since this is the conventional way of presenting partitioning data in the literature.

Finally, we will use our new partitioning values to calculate equilibrium liquids and compare those liquid compositions with analyses of natural CAIs to evaluate the degree to which those compositions might represent equilibrium.

### 5.1. Site occupancy and temperature/bulk composition dependence

**Grossite** ($CaAl_4O_7$): Grossite is a rare phase in terrestrial rocks, however, in extraterrestrial materials it is found in CV3, CO3 and CM2 chondrite CAIs as an accessory phase or rims around melilite, and common in CAIs in ALH 85085 (CH), Acfer 182 (CH) and El Djouf 001 (CR2) (Weber and Bischoff, 1994), and in CO and CM chondrites (Kööp et al., 2018; Simon et al., 2019a; Simon et al., 2019b).

Stoichiometrically, pure grossite has one octahedral site occupied by Ca, and 4 tetrahedral sites occupied by Al. Oxygen atoms form bridging layers of interlinked $AlO_4$ tetrahedra located in the same plane as distorted octahedral sites occupied by Ca (Karmaoui et al., 2009). In contrast to $SrAl_4O_7$, $CaAl_4O_7$ does not have an $AlO_6$ octahedron in addition to the $AlO_4$ tetrahedral layer. Al and Ca occupy different sites, and other large +3 and +2 cations will occupy the distorted octahedral site, and smaller +3 and +2 cations will occupy tetrahedral sites. It is also assumed that +2 cations will occupy the octahedral site with Ca. We deem this to be more likely than their preference for a tetrahedral site. However, that assumption should be tested in future research before we develop a predictive model or attempt to calculate elastic strain parameters.

All trace elements measured as part of this investigation are moderately to highly incompatible in grossite (Table 2). The light elements Li, Be and B are moderately incompatible, with Li and Be being slightly more compatible than B. The low value for B suggests that the tetrahedral sites normally occupied by Al are relatively inflexible. For the +2 cations, we assume that they occupy the same site occupied by Ca. The +2 cations exhibit a curve with an $r_o$ at ~1.1 Å. This is similar to the $r_o$ exhibited for the +3 cations where the curve is also between $Ce^{+3}$ and $Eu^{+3}$ with an $r_o$ close to 1.1 Å (Fig. 8). Eu is slightly more compatible than the other rare earth elements, suggesting the possibility that a fraction of Eu is present as $Eu^{+2}$, and thus behaves more like Sr and Ca and is therefore more compatible than the +3 REEs. However, this would require additional experiments to confirm. The +4 and +5 (Ti, Zr, Hf, Th, Nb and Ta) cations are generally more incompatible than the +2 and +3 cations. Given the magnitude of the analytical/experimental error for those elements, we cannot definitively determine the substitution mechanism for those elements with this set of experiments.

In an attempt to evaluate the magnitude of possible temperature dependence for $^{Grossite-Melt}D_i$ we compared the results of experiments #18 and #7 (Fig. 7). The major element compositions of grossite and liquid are essentially identical between these two experiments (Table 2). Experiment #18 is 60°C higher yet exhibits roughly similar partition coefficients for most of the trace elements measured compared to #7. Our contribution here is to present the first experimental partitioning data for grossite, and to document that the partition coefficients for the +2 and +3 cations exhibit similar $r_o$, and the +4 and +5 cations are sufficiently incompatible that we need to conduct further experimental work to obtain bigger crystals, and analye using instruments with smaller analytical volumes and lower detection limits.

**Table 2.** Major (wt%) and trace element composition (ppm) of grossite and melt pairs from experiments #18 and #7 with average grossite-melt partition coefficients. Detection limits (ppm) for the measured trace elements are given on the right. Bdl: below detection limit. Numbers within the parentheses are errors (1σ).

| Exp #18 | Pair 1 | | Pair 2 | | Pair 3 | | Pair 4 | | | | | | |
|---|---|---|---|---|---|---|---|---|---|---|---|---|---|
| Phase | grossite (#9) | melt (#5) | grossite (#9) | melt (#4) | grossite (#9) | melt (#6) | grossite (#7) | melt (#10) | | | | | |
| SiO$_2$ (wt%) | 0.13 *(0.03)* | 8.02 *(0.06)* | 0.13 *(0.03)* | 7.89 *(0.06)* | 0.13 *(0.03)* | 7.86 *(0.06)* | 0.12 *(0.03)* | 7.91 *(0.06)* | $^{Gr-Melt}D$ (averages from pairs) | | | | |
| TiO$_2$ | 0.02 *(0.01)* | 2.72 *(0.02)* | 0.02 *(0.01)* | 2.73 *(0.02)* | 0.02 *(0.01)* | 2.74 *(0.02)* | 0.01 *(0.01)* | 2.69 *(0.02)* | Li | bdl | | | |
| Al$_2$O$_3$ | 77.15 *(0.18)* | 49.66 *(0.17)* | 77.15 *(0.18)* | 50.16 *(0.17)* | 77.15 *(0.18)* | 50.77 *(0.17)* | 77.49 *(0.18)* | 50.06 *(0.17)* | Be | bdl | | ~ Detection limits | |
| FeO | 0.01 *(0.00)* | 0.06 *(0.01)* | 0.01 *(0.00)* | 0.08 *(0.01)* | 0.01 *(0.00)* | 0.07 *(0.01)* | 0.01 *(0.00)* | 0.06 *(0.01)* | B | bdl | | grossite | melt |
| MgO | 0.07 *(0.02)* | 0.45 *(0.01)* | 0.07 *(0.02)* | 0.46 *(0.01)* | 0.07 *(0.02)* | 0.43 *(0.01)* | 0.08 *(0.02)* | 0.44 *(0.01)* | Mg | 0.16 *(0.02)* | Li (ppm) | 2.96 | 3.05 |
| CaO | 22.30 *(0.05)* | 36.57 *(0.16)* | 22.30 *(0.05)* | 37.39 *(0.16)* | 22.30 *(0.05)* | 36.64 *(0.16)* | 22.19 *(0.05)* | 36.37 *(0.16)* | Ti | 0.01 *(0.01)* | Be | 2.20 | 1.38 |
| Cr$_2$O$_3$ | 0.00 *(0.00)* | 0.00 *(0.00)* | 0.00 *(0.00)* | 0.00 *(0.00)* | 0.00 *(0.00)* | 0.00 *(0.00)* | 0.00 *(0.00)* | 0.01 *(0.00)* | Ca | 0.61 *(0.03)* | B | 11.88 | 10.88 |
| Li (ppm) | bdl | 5.39 *(2.21)* | bdl | bdl | bdl | 4.72 *(1.37)* | bdl | 4.57 *(1.59)* | Sr | 0.29 *(0.02)* | Sr | 0.95 | 0.83 |
| Be | bdl | 2.54 *(0.74)* | bdl | bdl | bdl | 2.52 *(0.85)* | bdl | 2.53 *(1.07)* | Zr | bdl | Zr | 2.50 | 1.87 |
| B | bdl | 39.11 *(9.19)* | bdl | bdl | bdl | 35.28 *(5.07)* | bdl | 33.28 *(5.23)* | Nb | 0.03 *(0.01)* | Nb | 0.37 | 0.30 |
| Sr | 38.77 *(1.14)* | 133.49 *(0.90)* | 38.77 *(1.14)* | 134.23 *(2.05)* | 38.77 *(1.14)* | 123.74 *(0.62)* | 36.48 *(1.09)* | 131.86 *(0.71)* | Ba | bdl | Ba | 5.36 | 4.62 |
| Zr | bdl | 45.48 *(1.53)* | bdl | 48.98 *(4.19)* | bdl | 42.45 *(0.97)* | bdl | 45.17 *(1.10)* | La | 0.07 *(0.01)* | La | 0.36 | 0.33 |
| Nb | 1.89 *(0.40)* | 50.09 *(0.40)* | 1.89 *(0.40)* | 52.08 *(0.85)* | 1.89 *(0.40)* | 46.45 *(0.33)* | bdl | 50.49 *(0.32)* | Ce | 0.09 *(0.02)* | Ce | 0.20 | 0.16 |
| Ba | bdl | 77.89 *(4.01)* | bdl | 78.04 *(10.02)* | bdl | 76.81 *(2.39)* | bdl | 71.64 *(2.71)* | Eu | 0.10 *(0.01)* | Eu | 0.31 | 0.31 |
| La | 4.67 *(0.43)* | 68.28 *(0.46)* | 4.67 *(0.43)* | 70.94 *(1.00)* | 4.67 *(0.43)* | 60.27 *(0.35)* | 3.89 *(0.36)* | 68.79 *(0.39)* | Dy | 0.05 *(0.01)* | Dy | 0.19 | 0.33 |
| Ce | 5.88 *(0.28)* | 62.72 *(0.39)* | 5.88 *(0.28)* | 64.44 *(0.70)* | 5.88 *(0.28)* | 56.54 *(0.29)* | 3.77 *(0.26)* | 64.36 *(0.33)* | Ho | 0.04 *(0.01)* | Ho | 0.09 | 0.10 |
| Eu | 5.78 *(0.39)* | 57.62 *(0.51)* | 5.78 *(0.39)* | 61.18 *(1.03)* | 5.78 *(0.39)* | 51.75 *(0.35)* | 5.54 *(0.42)* | 58.93 *(0.39)* | Yb | 0.02 *(0.01)* | Yb | 0.35 | 0.18 |
| Dy | 2.23 *(0.29)* | 46.62 *(0.59)* | 2.23 *(0.29)* | 48.87 *(1.37)* | 2.23 *(0.29)* | 41.16 *(0.43)* | 2.05 *(0.34)* | 45.70 *(0.50)* | Hf | 0.003 | Hf | 0.17 | 0.15 |
| Ho | 1.68 *(0.14)* | 44.56 *(0.31)* | 1.68 *(0.14)* | 47.55 *(0.59)* | 1.68 *(0.14)* | 39.38 *(0.23)* | 1.63 *(0.12)* | 43.79 *(0.26)* | Ta | 0.02 *(0.01)* | Ta | 0.10 | 0.04 |
| Yb | 0.90 *(0.40)* | 44.42 *(0.63)* | 0.90 *(0.40)* | 48.57 *(1.09)* | 0.90 *(0.40)* | 39.33 *(0.45)* | 0.67 *(0.36)* | 45.14 *(0.55)* | Th | 0.02 *(0.01)* | Th | 0.02 | 0.01 |
| Hf | bdl | 38.35 *(0.55)* | bdl | 40.09 *(0.88)* | bdl | 34.25 *(0.39)* | 0.13 *(0.05)* | 38.23 *(0.43)* | | | | | |
| Ta | 0.29 *(0.12)* | 42.01 *(0.28)* | 0.29 *(0.12)* | 44.24 *(0.49)* | 0.29 *(0.12)* | 36.20 *(0.21)* | 2.45 *(0.19)* | 42.55 *(0.25)* | | | | | |
| Th | 0.22 *(0.05)* | 41.40 *(0.34)* | 0.22 *(0.05)* | 45.24 *(0.59)* | 0.22 *(0.05)* | 35.91 *(0.25)* | 2.16 *(0.22)* | 41.60 *(0.29)* | | | | | |

| Exp #7 | Pair 1 | | Pair 2 | | | | | | |
|---|---|---|---|---|---|---|---|---|---|
| Phase | grossite (#6) | melt (#8) | grossite (#6) | melt (#9) | | | | | |
| SiO$_2$ (wt%) | 0.33 *(0.06)* | 11.22 *(1.37)* | 0.33 *(0.06)* | 11.23 *(1.37)* | $^{Gr-Melt}D$ (averages from pairs) | | | | |
| TiO$_2$ | 0.01 | 3.23 *(0.71)* | 0.01 | 3.27 *(0.71)* | Li | bdl | | | |
| Al$_2$O$_3$ | 71.87 *(0.55)* | 46.22 *(1.19)* | 71.87 *(0.55)* | 46.85 *(1.19)* | Be | 0.43 *(0.01)* | | ~ Detection limits | |
| FeO | 0.00 | 0.00 | 0.00 | 0.00 | B | bdl | | grossite | melt |
| MgO | 0.11 *(0.07)* | 0.51 *(0.08)* | 0.11 *(0.07)* | 0.47 *(0.08)* | Mg | 0.22 *(0.03)* | Li (ppm) | 3.31 | 2.02 |
| CaO | 21.69 *(0.07)* | 33.76 *(0.50)* | 21.69 *(0.07)* | 33.90 *(0.50)* | Ti | 0.01 *(0.01)* | Be | 1.00 | 0.91 |
| Cr$_2$O$_3$ | 0.00 | 0.00 | 0.00 | 0.00 | Ca | 0.64 *(0.04)* | B | 5.39 | 3.55 |
| Li (ppm) | bdl | 9.12 *(2.02)* | bdl | 8.50 *(1.83)* | Sr | 0.31 *(0.01)* | Sr | 0.26 | 0.18 |
| Be | 106.50 *(2.66)* | 250.23 *(2.85)* | 106.50 *(2.66)* | 242.53 *(2.65)* | Zr | 0.09 *(0.01)* | Zr | 0.37 | 0.23 |
| B | bdl | 108.66 *(3.67)* | bdl | 105.14 *(4.01)* | Nb | 0.01 *(0.01)* | Nb | 0.07 | 0.05 |
| Sr | 55.79 *(0.41)* | 179.56 *(0.49)* | 55.79 *(0.41)* | 180.11 *(0.47)* | Ba | 0.03 *(0.01)* | Ba | 1.95 | 1.20 |
| Zr | 1.35 *(0.38)* | 168.89 *(0.57)* | 1.35 *(0.38)* | 169.57 *(0.54)* | La | 0.06 *(0.01)* | La | 0.11 | 0.08 |
| Nb | 1.26 *(0.08)* | 141.51 *(0.37)* | 1.26 *(0.08)* | 141.68 *(0.36)* | Ce | 0.07 *(0.01)* | Ce | 0.09 | 0.05 |
| Ba | 12.01 *(1.98)* | 465.56 *(2.00)* | 12.01 *(1.98)* | 466.45 *(1.89)* | Eu | 0.13 *(0.01)* | Eu | 0.07 | 0.06 |
| La | 3.04 *(0.13)* | 49.92 *(0.20)* | 3.04 *(0.13)* | 50.63 *(0.19)* | Dy | 0.04 *(0.01)* | Dy | 0.05 | 0.04 |
| Ce | 3.01 *(0.10)* | 45.95 *(0.17)* | 3.01 *(0.10)* | 46.00 *(0.17)* | Ho | 0.04 *(0.01)* | Ho | 0.02 | 0.02 |
| Eu | 8.42 *(0.13)* | 63.07 *(0.24)* | 8.42 *(0.13)* | 64.06 *(0.23)* | Yb | 0.03 *(0.01)* | Yb | 0.07 | 0.06 |
| Dy | 2.09 *(0.09)* | 50.29 *(0.28)* | 2.09 *(0.09)* | 50.89 *(0.27)* | Hf | 0.01 *(0.01)* | Hf | 0.07 | 0.04 |
| Ho | 2.34 *(0.05)* | 61.37 *(0.19)* | 2.34 *(0.05)* | 61.95 *(0.18)* | Ta | 0.01 *(0.01)* | Ta | 0.01 | 0.01 |
| Yb | 2.41 *(0.12)* | 85.53 *(0.40)* | 2.41 *(0.12)* | 87.68 *(0.39)* | Th | 0.01 *(0.01)* | Th | 0.01 | 0.01 |
| Hf | 1.16 *(0.09)* | 155.27 *(0.50)* | 1.16 *(0.09)* | 158.41 *(0.50)* | | | | | |
| Ta | 2.74 *(0.07)* | 336.75 *(0.77)* | 2.74 *(0.07)* | 340.34 *(0.75)* | | | | | |
| Th | 0.79 *(0.03)* | 75.37 *(0.25)* | 0.79 *(0.03)* | 77.38 *(0.24)* | | | | | |

**Table 3.** Major (wt%) and trace element composition (ppm) of gehlenite and melt pairs from experiments #6 and #7 with average gehlenite-melt partition coefficients. Detection limits (ppm) for the measured trace elements are given on the right. Bdl: below detection limit. Numbers within the parentheses are errors (1σ).

| Exp #6 | Pair 1 | | Pair 2 | | Pair 3 | | Pair 4 | |
|---|---|---|---|---|---|---|---|---|
| Phase | gehlenite (#10) | melt (#7) | gehlenite (#11) | melt (#7) | gehlenite (#11) | melt (#8) | gehlenite (#12) | melt (#8) |
| $SiO_2$ (wt%) | 20.21 (0.05) | 20.91 (0.14) | 21.44 (0.05) | 20.91 (0.14) | 21.44 (0.05) | 20.74 (0.14) | 21.38 (0.05) | 20.74 (0.14) |
| $TiO_2$ | 0.09 (0.04) | 2.38 (0.01) | 0.10 (0.04) | 2.38 (0.01) | 0.10 (0.04) | 2.40 (0.01) | 0.09 (0.04) | 2.40 (0.01) |
| $Al_2O_3$ | 38.43 (0.13) | 42.88 (0.04) | 38.24 (0.13) | 42.88 (0.04) | 38.24 (0.13) | 42.91 (0.04) | 37.53 (0.13) | 42.91 (0.04) |
| FeO | 0.00 | 0.04 (0.00) | 0.00 | 0.04 (0.00) | 0.00 | 0.04 (0.00) | 0.01 (0.00) | 0.04 (0.00) |
| MgO | 0.29 (0.01) | 0.51 (0.12) | 0.12 (0.01) | 0.51 (0.12) | 0.12 (0.01) | 0.51 (0.12) | 0.13 (0.01) | 0.51 (0.12) |
| CaO | 40.96 (0.12) | 32.66 (0.39) | 40.79 (0.12) | 32.66 (0.39) | 40.79 (0.12) | 32.66 (0.39) | 41.30 (0.12) | 32.66 (0.39) |
| $Cr_2O_3$ | 0.00 | 0.00 | 0.00 | 0.00 | 0.00 | 0.00 | 0.00 | 0.00 |
| Li (ppm) | bdl | 38.45 (2.59) | bdl | 38.45 (2.59) | bdl | 30.58 (2.43) | bdl | 30.58 (2.43) |
| Be | bdl | 6.39 (0.68) | 0.94 (0.52) | 6.39 (0.68) | 0.94 (0.52) | 5.91 (0.52) | bdl | 5.91 (0.52) |
| B | 5.32 (4.94) | 93.82 (6.04) | 6.92 (4.72) | 93.82 (6.04) | 6.92 (4.72) | 73.93 (4.59) | 5.42 (4.90) | 73.93 (4.59) |
| Sr | 123.06 (0.46) | 164.53 (0.54) | 121.92 (0.86) | 164.53 (0.54) | 121.92 (0.86) | 131.37 (0.43) | 120.38 (0.37) | 131.37 (0.43) |
| Zr | 1.35 (0.18) | 109.65 (0.42) | 1.50 (0.16) | 109.65 (0.42) | 1.50 (0.16) | 80.71 (0.31) | 1.58 (0.20) | 80.71 (0.31) |
| Nb | 1.08 (0.04) | 126.97 (0.32) | 1.58 (0.05) | 126.97 (0.32) | 1.58 (0.05) | 100.97 (0.24) | 1.78 (0.05) | 100.97 (0.24) |
| Ba | 10.06 (1.26) | 239.32 (1.67) | 12.12 (1.14) | 239.32 (1.67) | 12.12 (1.14) | 193.59 (1.40) | 12.42 (1.05) | 193.59 (1.40) |
| La | 65.80 (0.18) | 64.40 (0.21) | 58.77 (0.20) | 64.40 (0.21) | 58.77 (0.20) | 50.55 (0.16) | 59.92 (0.17) | 50.55 (0.16) |
| Ce | 78.07 (0.19) | 66.34 (0.20) | 72.78 (0.38) | 66.34 (0.20) | 72.78 (0.38) | 53.58 (0.15) | 74.50 (0.18) | 53.58 (0.15) |
| Eu | 88.12 (0.28) | 79.51 (0.30) | 88.86 (0.28) | 79.51 (0.30) | 88.86 (0.28) | 63.23 (0.24) | 87.51 (0.26) | 63.23 (0.24) |
| Dy | 61.96 (0.25) | 59.47 (0.28) | 61.65 (0.28) | 59.47 (0.28) | 61.65 (0.28) | 44.62 (0.21) | 59.58 (0.25) | 44.62 (0.21) |
| Ho | 71.06 (0.16) | 75.76 (0.20) | 71.31 (0.16) | 75.76 (0.20) | 71.31 (0.16) | 56.10 (0.14) | 67.77 (0.15) | 56.10 (0.14) |
| Yb | 55.10 (0.25) | 86.70 (0.38) | 56.04 (0.25) | 86.70 (0.38) | 56.04 (0.25) | 65.18 (0.27) | 52.42 (0.24) | 65.18 (0.27) |
| Hf | 2.42 (0.06) | 97.36 (0.35) | 1.53 (0.05) | 97.36 (0.35) | 1.53 (0.05) | 71.74 (0.25) | 1.71 (0.05) | 71.74 (0.25) |
| Ta | 1.50 (0.02) | 107.78 (0.25) | 1.49 (0.02) | 107.78 (0.25) | 1.49 (0.02) | 83.33 (0.18) | 1.62 (0.02) | 83.33 (0.18) |
| Th | 21.85 (0.10) | 70.48 (0.22) | 19.79 (0.09) | 70.48 (0.22) | 19.79 (0.09) | 54.54 (0.16) | 18.70 (0.09) | 54.54 (0.16) |

| Exp #6 | Pair 5 | | Pair 6 | | Pair 7 | | $^{Geh-Melt}D$ (averages from pairs) | | | ~ Detection limits | |
|---|---|---|---|---|---|---|---|---|---|---|---|
| Phase | gehlenite (#10) | melt (#9) | gehlenite (#11) | melt (#9) | gehlenite (#12) | melt (#9) | | | | gehlenite | melt |
| $SiO_2$ (wt%) | 20.21 (0.05) | 20.86 (0.14) | 21.44 (0.05) | 20.86 (0.14) | 21.38 (0.05) | 20.86 (0.14) | | | | | |
| $TiO_2$ | 0.09 (0.04) | 2.40 (0.01) | 0.10 (0.04) | 2.40 (0.01) | 0.09 (0.04) | 2.40 (0.01) | Li | bdl | | | |
| $Al_2O_3$ | 38.43 (0.13) | 42.91 (0.04) | 38.24 (0.13) | 42.91 (0.04) | 37.53 (0.13) | 42.91 (0.04) | Be | 0.14 (0.01) | | | |
| FeO | 0.00 | 0.04 (0.00) | 0.00 | 0.04 (0.00) | 0.01 (0.00) | 0.04 (0.00) | B | 0.07 (0.01) | | | |
| MgO | 0.29 (0.01) | 0.51 (0.12) | 0.12 (0.01) | 0.51 (0.12) | 0.13 (0.01) | 0.51 (0.12) | Mg | 0.14 (0.02) | Li (ppm) | 2.39 | 2.53 |
| CaO | 40.96 (0.12) | 32.66 (0.39) | 40.79 (0.12) | 32.66 (0.39) | 41.30 (0.12) | 32.66 (0.39) | Ti | 0.14 (0.03) | Be | 0.49 | 0.61 |
| $Cr_2O_3$ | 0.00 | 0.00 | 0.00 | 0.00 | 0.00 | 0.00 | Ca | 1.25 (0.04) | B | 4.83 | 5.37 |
| Li (ppm) | bdl | 37.59 (2.66) | bdl | 37.59 (4.32) | bdl | 37.59 (2.66) | Sr | 0.79 (0.09) | Sr | 0.52 | 0.38 |
| Be | bdl | 7.24 (0.89) | 0.94 (0.52) | 7.24 (0.67) | bdl | 7.24 (0.89) | Zr | 0.02 (0.00) | Zr | 0.18 | 0.20 |
| B | 5.32 (4.94) | 94.97 (5.85) | 6.92 (4.72) | 94.77 (11.83) | 5.42 (4.90) | 94.97 (5.85) | Nb | 0.01 (0.00) | Nb | 0.04 | 0.05 |
| Sr | 123.06 (0.46) | 165.30 (0.58) | 121.92 (0.86) | 165.30 (19.37) | 120.38 (0.37) | 165.30 (0.58) | Ba | 0.05 (0.01) | Ba | 1.13 | 1.32 |
| Zr | 1.35 (0.18) | 104.03 (0.42) | 1.50 (0.16) | 104.03 (15.34) | 1.58 (0.20) | 104.03 (0.42) | La | 1.03 (0.11) | La | 0.09 | 0.07 |
| Nb | 1.08 (0.04) | 127.61 (0.32) | 1.58 (0.05) | 127.61 (15.20) | 1.78 (0.05) | 127.61 (0.32) | Ce | 1.20 (0.13) | Ce | 0.14 | 0.07 |
| Ba | 10.06 (1.26) | 244.57 (1.69) | 12.12 (1.14) | 244.57 (28.04) | 12.42 (1.05) | 244.57 (1.69) | Eu | 1.19 (0.14) | Eu | 0.17 | 0.19 |
| La | 65.80 (0.18) | 63.58 (0.20) | 58.77 (0.20) | 63.58 (7.77) | 59.92 (0.17) | 63.58 (0.20) | Dy | 1.14 (0.15) | Dy | 0.08 | 0.05 |
| Ce | 78.07 (0.19) | 66.95 (0.19) | 72.78 (0.38) | 66.95 (7.55) | 74.50 (0.18) | 66.95 (0.19) | Ho | 1.04 (0.14) | Ho | 0.02 | 0.07 |
| Eu | 88.12 (0.28) | 79.60 (0.33) | 88.86 (0.28) | 79.60 (9.43) | 87.51 (0.26) | 79.60 (0.33) | Yb | 0.70 (0.09) | Yb | 0.06 | 0.10 |
| Dy | 61.96 (0.25) | 57.56 (0.27) | 61.65 (0.28) | 57.56 (8.08) | 59.58 (0.25) | 57.56 (0.27) | Hf | 0.02 (0.01) | Hf | 0.04 | 0.08 |
| Ho | 71.06 (0.16) | 72.23 (0.24) | 71.31 (0.16) | 72.23 (10.48) | 67.77 (0.15) | 72.23 (0.24) | Ta | 0.02 (0.01) | Ta | 0.01 | 0.07 |
| Yb | 55.10 (0.25) | 83.75 (0.37) | 56.04 (0.25) | 83.75 (11.67) | 52.42 (0.24) | 83.75 (0.37) | Th | 0.31 (0.03) | Th | 0.01 | 0.05 |
| Hf | 2.42 (0.06) | 93.75 (0.36) | 1.53 (0.05) | 93.75 (13.86) | 1.71 (0.05) | 93.75 (0.36) | | | | | |
| Ta | 1.50 (0.02) | 106.89 (0.27) | 1.49 (0.02) | 106.89 (13.86) | 1.62 (0.02) | 106.89 (0.27) | | | | | |
| Th | 21.85 (0.10) | 69.15 (0.22) | 19.79 (0.09) | 69.15 (8.84) | 18.70 (0.09) | 69.15 (0.22) | | | | | |

**Table 3.** (continued)

| Exp #7 | Pair 1 | | Pair 2 | | | | | | | |
|---|---|---|---|---|---|---|---|---|---|---|
| Phase | gehlenite (#7) | melt (#8) | gehlenite (#7) | melt (#9) | | | | | | |
| SiO$_2$ (wt%) | 20.82 *(3.32)* | 11.22 *(1.37)* | 20.82 *(3.32)* | 11.23 *(1.37)* | | | | | | |
| TiO$_2$ | 0.26 *(0.01)* | 3.23 *(0.71)* | 0.26 *(0.01)* | 3.27 *(0.71)* | $^{Geh-Melt}D$ (averages from pairs) | | | | | |
| Al$_2$O$_3$ | 34.50 *(3.66)* | 46.22 *(1.19)* | 34.50 *(3.66)* | 46.85 *(1.19)* | Li | bdl | | ~ Detection limits | | |
| FeO | 0.00 | 0.00 | 0.00 | 0.00 | Be | bdl | | gehlenite | melt | |
| MgO | 0.03 *(0.00)* | 0.51 *(0.08)* | 0.03 *(0.00)* | 0.47 *(0.08)* | B | bdl | Li (ppm) | 6.27 | 2.02 | |
| CaO | 41.22 *(2.64)* | 33.76 *(0.50)* | 41.22 *(2.64)* | 33.90 *(0.50)* | Mg | 0.06 *(0.02)* | Be | 5.24 | 0.91 | |
| Cr$_2$O$_3$ | 0.00 | 0.00 | 0.00 | 0.00 | Ti | 0.08 *(0.03)* | B | 9.18 | 3.55 | |
| Li (ppm) | bdl | 9.12 *(2.00)* | bdl | 8.50 *(1.81)* | Ca | 1.22 *(0.03)* | Sr | 0.57 | 0.18 | |
| Be | bdl | 250.23 *(1.21)* | bdl | 242.53 *(0.99)* | Sr | 0.79 *(0.02)* | Zr | 0.83 | 0.23 | |
| B | bdl | 108.66 *(3.36)* | bdl | 105.14 *(3.76)* | Zr | 0.02 *(0.01)* | Nb | 0.18 | 0.05 | |
| Sr | 141.74 *(0.57)* | 179.56 *(0.18)* | 141.74 *(0.57)* | 180.11 *(0.15)* | Nb | 0.02 *(0.01)* | Ba | 3.64 | 1.20 | |
| Zr | 3.73 *(0.83)* | 168.89 *(0.26)* | 3.73 *(0.83)* | 169.57 *(0.21)* | Ba | 0.05 *(0.01)* | La | 0.24 | 0.08 | |
| Nb | 2.32 *(0.18)* | 141.51 *(0.04)* | 2.32 *(0.18)* | 141.68 *(0.04)* | La | 3.44 *(0.03)* | Ce | 0.15 | 0.05 | |
| Ba | 22.40 *(3.64)* | 465.56 *(1.24)* | 22.40 *(3.64)* | 466.45 *(1.10)* | Ce | 3.86 *(0.01)* | Eu | 0.15 | 0.06 | |
| La | 173.16 *(0.24)* | 49.92 *(0.08)* | 173.16 *(0.24)* | 50.63 *(0.08)* | Eu | 2.88 *(0.03)* | Dy | 0.11 | 0.04 | |
| Ce | 177.33 *(0.15)* | 45.95 *(0.05)* | 177.33 *(0.15)* | 46.00 *(0.04)* | Dy | 3.23 *(0.03)* | Ho | 0.05 | 0.02 | |
| Eu | 183.33 *(0.15)* | 63.07 *(0.06)* | 183.33 *(0.15)* | 64.06 *(0.05)* | Ho | 2.91 *(0.02)* | Yb | 0.16 | 0.06 | |
| Dy | 163.49 *(0.11)* | 50.29 *(0.03)* | 163.49 *(0.11)* | 50.89 *(0.05)* | Yb | 1.70 *(0.03)* | Hf | 0.14 | 0.04 | |
| Ho | 179.12 *(0.05)* | 61.37 *(0.02)* | 179.12 *(0.05)* | 61.95 *(0.02)* | Hf | 0.02 *(0.01)* | Ta | 0.02 | 0.01 | |
| Yb | 147.56 *(0.16)* | 85.53 *(0.06)* | 147.56 *(0.16)* | 87.68 *(0.03)* | Ta | 0.01 *(0.01)* | Th | 0.01 | 0.01 | |
| Hf | 3.23 *(0.14)* | 155.27 *(0.05)* | 3.23 *(0.14)* | 158.41 *(0.04)* | Th | 1.71 *(0.03)* | | | | |
| Ta | 4.50 *(0.13)* | 336.75 *(1.45)* | 4.50 *(0.13)* | 340.34 *(1.04)* | | | | | | |
| Th | 130.50 *(0.00)* | 75.37 *(0.01)* | 130.50 *(0.00)* | 77.38 *(0.01)* | | | | | | |

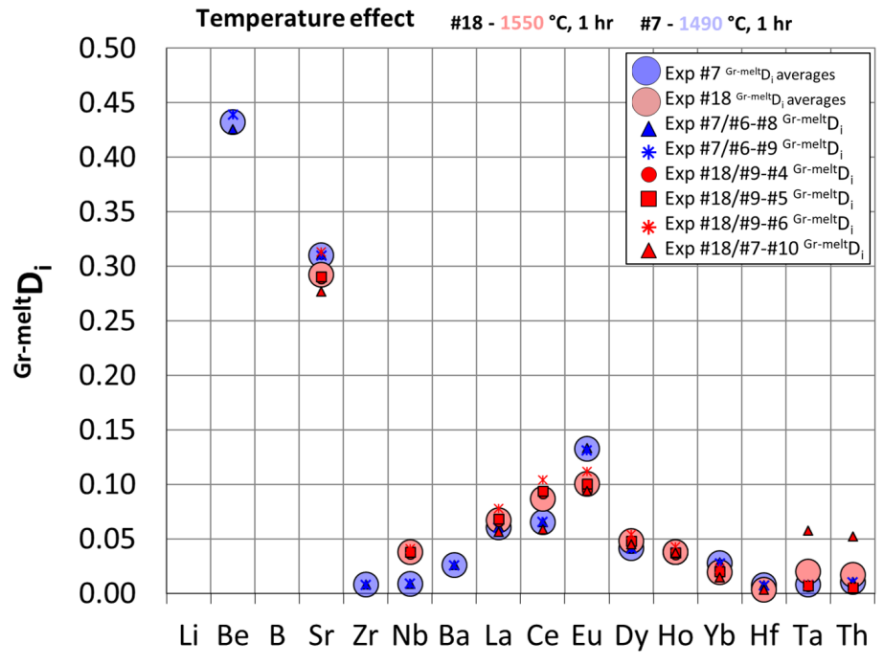

**Figure 7.** Temperature control on trace and rare earth element partitioning between grossite and melt ($^{Gr-melt}D_i$) for experiment #18 at 1550 °C, 1 hr; and #7 at 1490 °C, 1 hr. Large filled, light-colored circles show the average $^{Gr-melt}D_i$ for the given experiment. Small filled, dark colored symbols show the $^{Gr-melt}D_i$ from individual mineral-melt pair of the same colored experiment. Red indicates high temperature runs, blue indicates lower temperature.

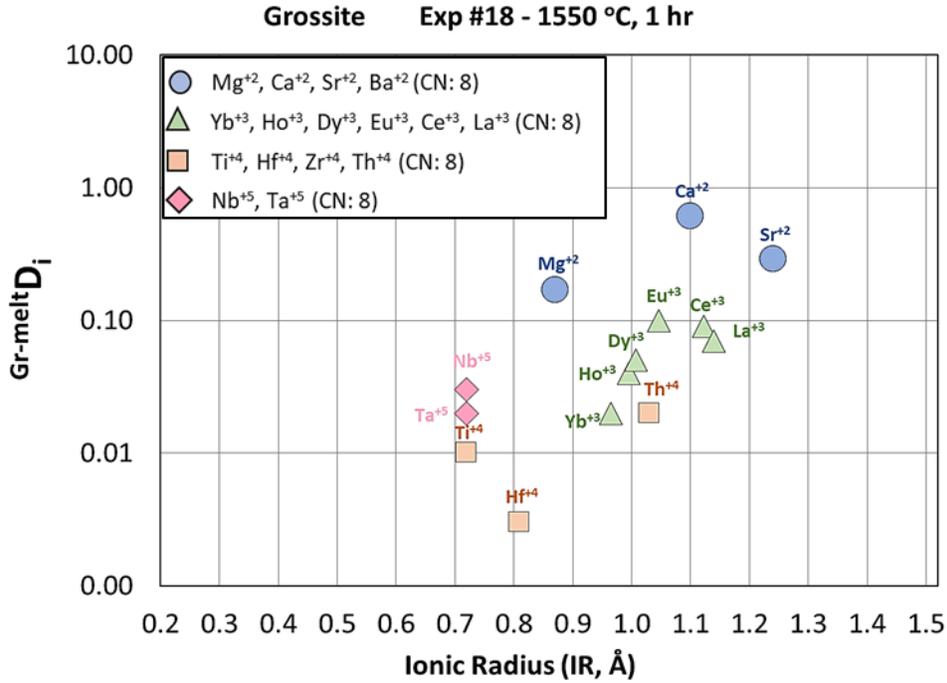

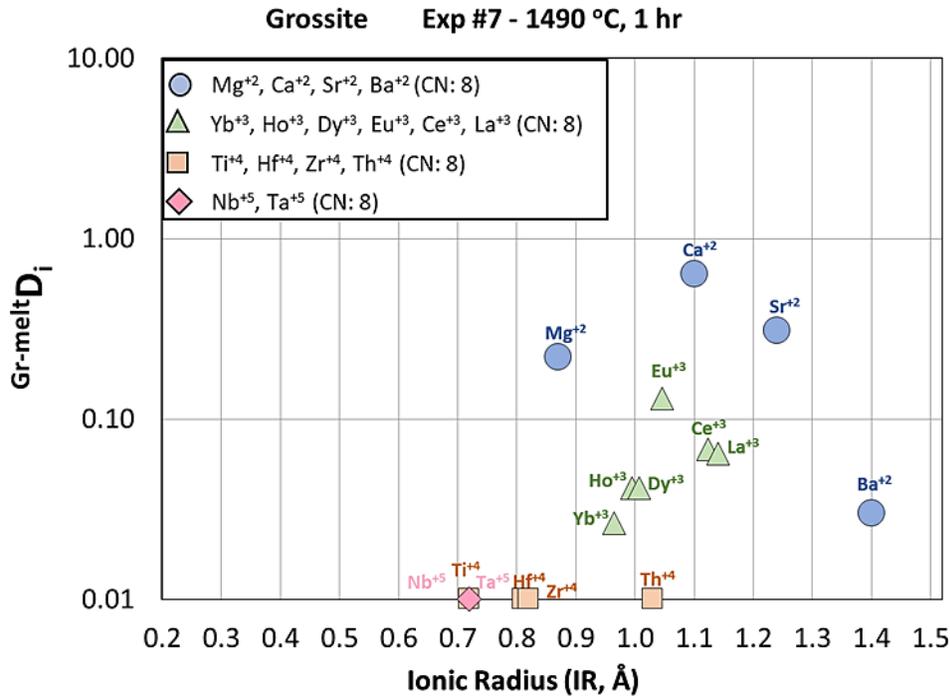

**Figure 8.** Onuma diagram for trace element partitioning between grossite and melt and $^{Gr\text{-}melt}D_i$ for two temperatures **a.** #18 at 1550 °C, 1 hr and **b.** #7 at 1490 °C, 1 hr. Ionic radii reflect values for coordination number (CN) = 8 cations (Shannon, 1976).

**Gehlenite** (**$Ca_2Al[AlSiO_7]$**): The melilite structure consists of sheets of tetrahedra containing Si, Al, and Mg (in the akermanite end-member). The sheets are bonded together by distorted eight-fold coordination sites normally occupied by Ca. There are three distinct cation sites: The X site, the eight-fold coordination site; the T1 site, a tetrahedral site usually occupied by Al or Mg; and a smaller tetrahedral (T2) site, usually occupied by Si or Al (Bindi et al., 2001).

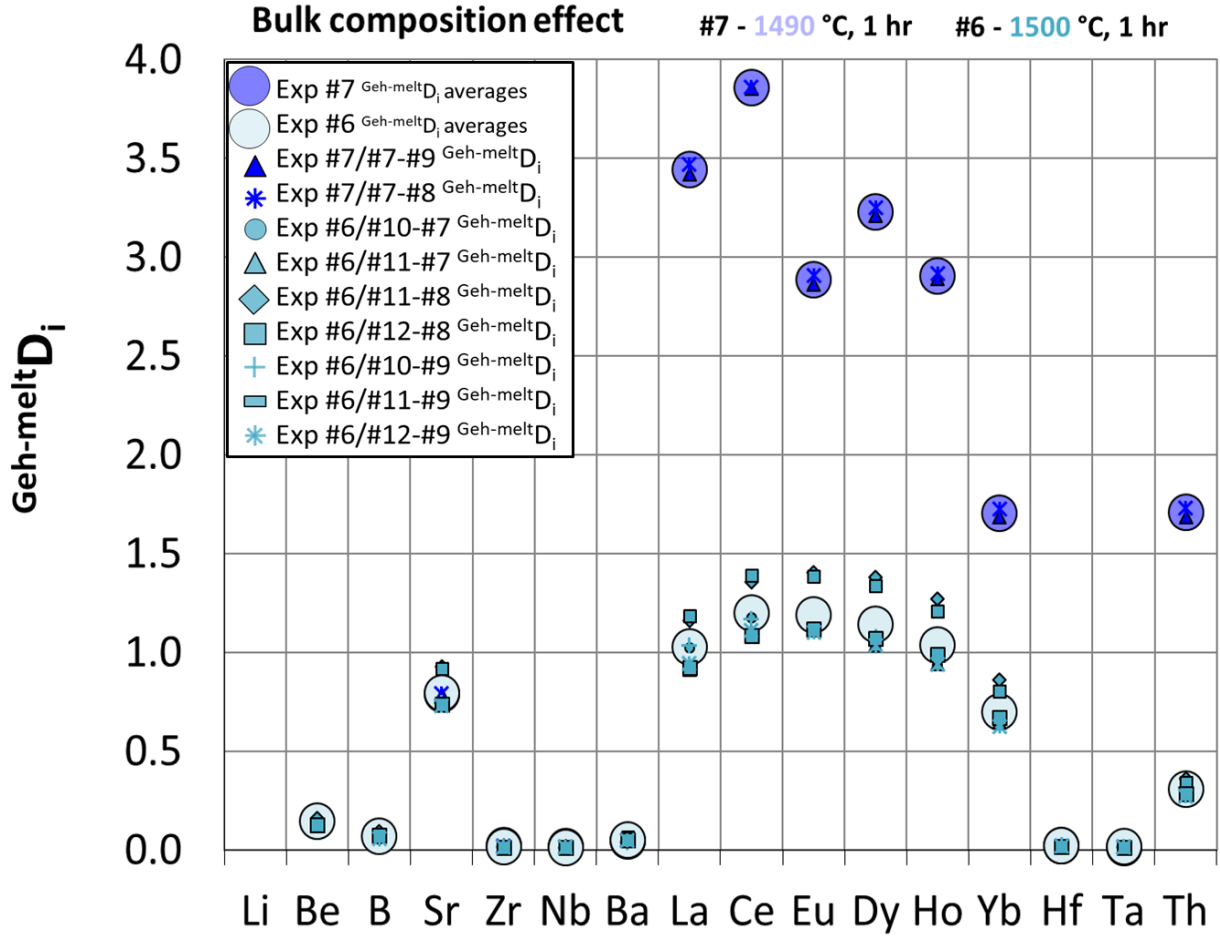

**Figure 9.** Bulk composition controls on trace and rare earth element partitioning between gehlenite and melt ($^{\text{Geh-melt}}D_i$) for experiments #7 at 1490 °C, 1 hr; and #6 at 1500 °C, 1 hr. Large filled and light colored circles show the average $^{\text{Geh-melt}}D_i$ for each experiment. Small filled and dark colored symbols show the $^{\text{Geh-melt}}D_i$ from individual mineral-melt pair of the same colored experiment.

The composition of the melilite in our experiments differs from the literature partitioning experiments in that our products were pure gehlenite. The rare earth elements (La, Ce, Eu, Dy, Ho, and Yb) are moderately incompatible to moderately compatible in gehlenite with partition coefficients ~ 1 (Table 3; Fig. 9). This is generally consistent with the trends documented in Beckett et al. (1990), whose results for starting compositions that were higher in Ca and lower in Al showed partitioning of the REE that were positively correlated to gehlenite content. The compatibilities of La, Ce, Eu, Dy, Ho in gehlenite ($Ak_0$) in our experiments are substantially higher than those of Lundstrom et al (2006) or Beckett et al. (1990). However, our results are consistent in that the temperature dependencies reported by Beckett et al. (1990) and Lundstrom et al. (2006) are similar to our results. In addition, our results are similar to those of Kuehner et al. (1989) for $Ak_{12}$ and Nagasawa et al. (1980) for an Ak-rich bulk composition (major element composition of the melilite was not reported). Our experiments were conducted at significantly higher temperature. However, isolating a temperature dependence from the compositional dependence is difficult because of the systematic covariation of temperature and composition in the systems on which we performed the experiments (Kuehner et al., 1989). As noted above, the development of a predictive model that takes the compositional and temperature dependence into account will require significantly more data. However, these new experiments will inform the boundary conditions for that future work.

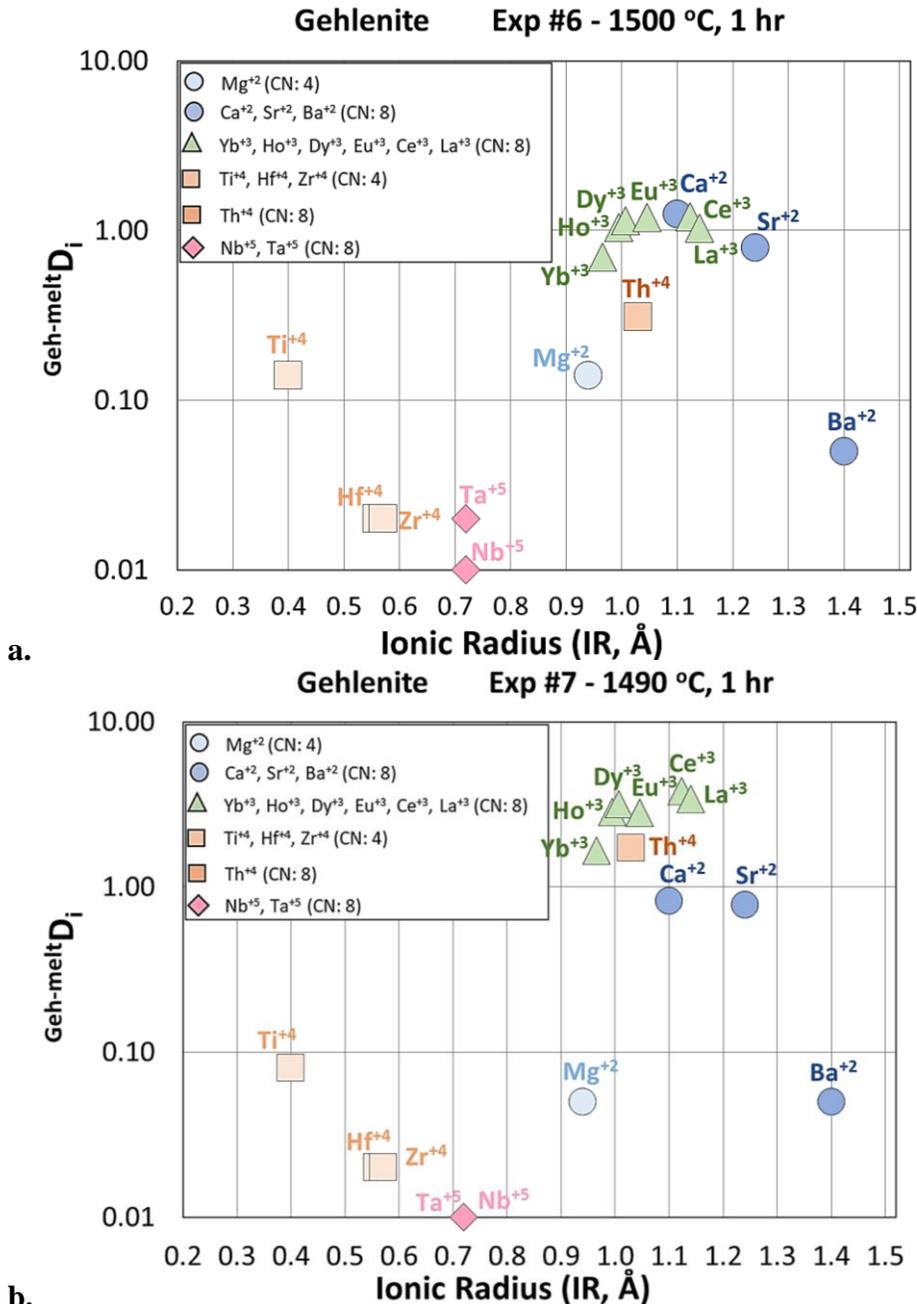

**Figure 10.** Onuma diagram for trace element partitioning between gehlenite and melt for two different bulk compositions showing $^{Geh\text{-}melt}D_i$ for experiments **a.** #6 at 1500 °C, 1 hr and **b.** #7 at 1490°C, 1 hr. Ionic radii are values for CN = 4 and CN = 8 cations (Shannon, 1976).

**Table 4.** Major (wt%) and trace element composition (ppm) of hibonite and melt pairs from experiments #6 and #7 with average hibonite-melt partition coefficients. Detection limits (ppm) for the measured trace elements are given on the right. Bdl: below detection limit. Numbers within the parentheses are errors (1σ).

| Exp #6 | Pair 1 | | Pair 2 | | | | | | |
|---|---|---|---|---|---|---|---|---|---|
| Phase | hibonite (#5) | melt (#7) | hibonite (#5) | melt (#8) | | | | | |
| $SiO_2$ (wt%) | 0.70 (0.04) | 20.91 (0.14) | 0.70 (0.04) | 20.74 (0.14) | | | | | |
| $TiO_2$ | 2.37 (0.04) | 2.38 (0.01) | 2.37 (0.04) | 2.40 (0.01) | $^{Hib-Melt}D$ (averages from pairs) | | | | |
| $Al_2O_3$ | 88.06 (1.85) | 42.88 (0.04) | 88.06 (1.85) | 42.91 (0.04) | Li | bdl | | | |
| FeO | 0.00 | 0.04 (0.00) | 0.00 | 0.04 (0.00) | Be | bdl | | ~ Detection limits | |
| MgO | 1.61 (0.03) | 0.51 (0.12) | 1.61 (0.03) | 0.51 (0.12) | B | bdl | | hibonite | melt |
| CaO | 8.47 (0.13) | 32.60 (0.39) | 8.47 (0.13) | 32.66 (0.39) | Mg | 1.07 (0.03) | Li (ppm) | 2.91 | 2.53 |
| $Cr_2O_3$ | 0.03 (0.01) | 0.00 | 0.03 (0.01) | 0.00 | Ti | 0.99 (0.03) | Be | 0.64 | 0.61 |
| Li (ppm) | bdl | 34.51 (2.51) | bdl | 30.58 (2.43) | Ca | 0.26 (0.04) | B | 7.18 | 5.37 |
| Be | bdl | 6.15 (0.60) | bdl | 5.91 (0.52) | Sr | 0.21 (0.02) | Sr | 0.44 | 0.38 |
| B | bdl | 83.87 (5.32) | bdl | 73.93 (4.59) | Zr | 0.02 (0.01) | Zr | 0.22 | 0.20 |
| Sr | 29.68 (0.50) | 147.95 (0.49) | 29.68 (0.50) | 131.37 (0.43) | Nb | 0.03 (0.01) | Nb | 0.05 | 0.05 |
| Zr | 1.82 (0.24) | 95.18 (0.37) | 1.82 (0.24) | 80.71 (0.31) | Ba | 0.02 (0.01) | Ba | 1.24 | 1.32 |
| Nb | 0.50 (0.06) | 113.97 (0.28) | 0.50 (0.06) | 100.97 (0.24) | La | 0.56 (0.05) | La | 0.08 | 0.07 |
| Ba | 3.52 (1.26) | 216.46 (1.53) | 3.52 (1.26) | 193.59 (1.40) | Ce | 0.52 (0.04) | Ce | 0.04 | 0.05 |
| La | 30.35 (0.24) | 57.47 (0.18) | 30.35 (0.24) | 50.55 (0.16) | Eu | 0.33 (0.03) | Eu | 0.17 | 0.19 |
| Ce | 29.18 (0.22) | 59.96 (0.17) | 29.18 (0.22) | 53.58 (0.15) | Dy | 0.32 (0.03) | Dy | 0.05 | 0.05 |
| Eu | 22.41 (0.26) | 71.37 (0.27) | 22.41 (0.26) | 63.23 (0.24) | Ho | 0.29 (0.03) | Ho | 0.01 | 0.07 |
| Dy | 15.21 (0.22) | 52.05 (0.25) | 15.21 (0.22) | 44.62 (0.21) | Yb | 0.19 (0.02) | Yb | 0.06 | 0.10 |
| Ho | 17.64 (0.14) | 65.93 (0.17) | 17.64 (0.14) | 56.10 (0.14) | Hf | 0.05 (0.01) | Hf | 0.04 | 0.08 |
| Yb | 13.34 (0.22) | 75.94 (0.33) | 13.34 (0.22) | 65.18 (0.27) | Ta | 0.02 (0.01) | Ta | 0.01 | 0.07 |
| Hf | 3.64 (0.10) | 84.55 (0.30) | 3.64 (0.10) | 71.74 (0.25) | Th | 0.19 (0.02) | Th | 0.01 | 0.05 |
| Ta | 1.51 (0.03) | 95.56 (0.22) | 1.51 (0.03) | 83.33 (0.18) | | | | | |
| Th | 11.27 (0.13) | 62.51 (0.19) | 11.27 (0.13) | 54.54 (0.16) | | | | | |
| | | | | | | | | | |
| Exp #7 | Pair 1 | | Pair 2 | | | | | | |
| Phase | hibonite (#10) | melt (#12) | hibonite (#10) | melt (#8,9) | | | | | |
| $SiO_2$ (wt%) | 0.22 (0.01) | 10.82 (1.37) | 0.22 (0.01) | 11.23 (1.37) | | | | | |
| $TiO_2$ | 1.54 (0.68) | 3.46 (0.71) | 1.54 (0.68) | 3.27 (0.71) | $^{Hib-Melt}D$ (averages from pairs) | | | | |
| $Al_2O_3$ | 91.05 (1.78) | 50.48 (1.19) | 91.05 (1.78) | 46.85 (1.19) | Li | bdl | | | |
| FeO | 0.00 | 0.00 | 0.00 | 0.00 | Be | 0.04 (0.01) | | ~ Detection limits | |
| MgO | 0.89 (0.05) | 0.49 (0.08) | 0.89 (0.05) | 0.47 (0.08) | B | bdl | | hibonite | melt |
| CaO | 8.64 (0.11) | 34.30 (0.50) | 8.64 (0.11) | 33.90 (0.50) | Mg | 1.85 (0.04) | Li (ppm) | 6.92 | 2.02 |
| $Cr_2O_3$ | 0.00 | 0.00 | 0.00 | 0.00 | Ti | 0.46 (0.03) | Be | 5.15 | 0.91 |
| Li (ppm) | bdl | 10.05 (2.27) | bdl | 8.81 (1.92) | Ca | 0.25 (0.03) | B | 10.40 | 3.55 |
| Be | 9.94 (5.25) | 263.58 (2.78) | 9.94 (5.25) | 246.38 (2.75) | Sr | 0.66 (0.03) | Sr | 0.65 | 0.18 |
| B | bdl | 116.48 (3.86) | bdl | 106.90 (3.84) | Zr | 0.18 (0.02) | Zr | 0.85 | 0.23 |
| Sr | 122.89 (1.10) | 192.66 (0.52) | 122.89 (1.10) | 179.83 (0.48) | Nb | 0.05 (0.01) | Nb | 0.14 | 0.05 |
| Zr | 31.60 (0.935) | 193.59 (0.61) | 31.60 (0.935) | 169.23 (0.55) | Ba | 0.06 (0.02) | Ba | 4.05 | 1.20 |
| Nb | 7.55 (0.18) | 163.33 (0.42) | 7.55 (0.18) | 141.59 (0.37) | La | 4.38 (0.18) | La | 0.29 | 0.08 |
| Ba | 27.44 (4.10) | 519.35 (2.14) | 27.44 (4.10) | 465.00 (1.94) | Ce | 3.54 (0.11) | Ce | 0.26 | 0.05 |
| La | 213.99 (1.45) | 47.45 (0.20) | 213.99 (1.45) | 50.28 (0.19) | Eu | 0.84 (0.04) | Eu | 0.16 | 0.06 |
| Ce | 159.25 (1.11) | 44.06 (0.17) | 159.25 (1.11) | 45.98 (0.17) | Dy | 0.25 (0.01) | Dy | 0.13 | 0.04 |
| Eu | 53.38 (0.50) | 63.34 (0.25) | 53.38 (0.50) | 63.57 (0.24) | Ho | 0.17 (0.02) | Ho | 0.06 | 0.02 |
| Dy | 12.34 (0.28) | 48.17 (0.28) | 12.34 (0.28) | 50.59 (0.28) | Yb | 0.05 (0.01) | Yb | 0.16 | 0.06 |
| Ho | 10.31 (0.14) | 61.07 (0.19) | 10.31 (0.14) | 61.66 (0.18) | Hf | 0.38 (0.03) | Hf | 0.15 | 0.04 |
| Yb | 4.04 (0.21) | 91.15 (0.43) | 4.04 (0.21) | 86.61 (0.40) | Ta | 0.17 (0.02) | Ta | 0.05 | 0.01 |
| Hf | 63.85 (0.64) | 177.98 (0.57) | 63.85 (0.64) | 156.84 (0.50) | Th | 0.81 (0.01) | Th | 0.08 | 0.01 |
| Ta | 62.77 (0.55) | 395.65 (0.89) | 62.77 (0.55) | 338.54 (0.76) | | | | | |
| Th | 62.77 (0.51) | 78.21 (0.26) | 62.77 (0.51) | 76.38 (0.24) | | | | | |

The compatibility of $La^{+3}$, $Ce^{+3}$, $Eu^{+3}$, $Dy^{+3}$, and $Ho^{+3}$ can be explained by these $REE^{+3}$ occupying the same site as $Ca^{+2}$ in the structure of gehlenite. The ideal radius for the Ca site is ~ 1.1 Å (Fig. 10) (Lundstrom et al., 2006), near the ionic radius of the middle REE. The absence of an Eu anomaly for these experiments, in spite of the low experimental $fO_2$ (IW+1), may be explained by the similarity of the partition coefficients for Eu +2 and +3 for gehlenite. This contention is supported by the similarity of the partitioning behavior for the MREE (e.g., Dy) and Sr for this composition (Table 3).

Based on their similarity in valence and ionic radius to Ca, the LFSEs ($Ba^{+2}$, $Sr^{+2}$) might be expected to have a relatively high $^{Gehlenite-Melt}D_i$ (Lundstrom et al., 2006; Kuehner et al., 1989) However, $Ba^{+2}$ was found to be incompatible in our experiments, consistent with the results of Lundstrom et al. (2006) and consistent with a narrow parabola on an Onuma diagram (Fig. 10) for +2 cations in eight-fold coordination as proposed by Lundstrom et al. (2006). In effect, the ideal radius is between Ca and Sr at ~1.1 Å and Ba at 1.4 Å is too large to fit.

The HFSEs, Ti, Zr, Nb, Ta and Hf are all incompatible between gehlenite and CAI-type liquids, with Ti being the most compatible. Existing data on melilite (Lundstrom et al., 2006, their Fig. 3) for $Ak_{30-70}$ in CAI-type melts showed significantly more incompatible behavior of Zr, Nb, Ta, Hf and Th. This suggests the possibility of a strong compositional dependence for at least Th and Ta partitioning between melilite and CAI-type liquids. Specifically, the smaller +4 and +5 cations (Ti, Hf Zr, Ta, Nb) exhibit a pattern consistent with substitution into a smaller tetrahedral site. However, Th is significantly larger, and under circumstances such as paired substitution with either vacancies or with Al, may fit into the same site as the REE or Ca. This is a potentially important issue with respect to our understanding of the behavior of Th in the early solar system but will require additional experiments to illuminate.

In our experiments, the light elements Li, Be, B are moderately to highly incompatible in gehlenite (Table 2). Boron and Be are expected to occupy the tetrahedral sites (Lundstrom et al., 2006). The site occupancy of Li is ambiguous, with an ionic radius between the ideal for the tetrahedral and octahedral sites. Beckett et.al. (1990) and Lundstrom et al. (2006) reported values for $^{Gehlenite-Melt}D_{Be}$ indicating that Be was compatible above $Ak_{40}$ at low temperatures. In addition, Becket et al., (2000) documented that Na was increasingly compatible as Ak increased. Future work should attempt to quantify the details of the correlation of the behavior of these +1 and +2 cations. However, existing data with our new data indicate increasing $D$s in the order $B^{+3}$> $Be^{+2}$>$Li^{+1}$ consistent with an interpretation that ionic radius (smaller size) has stronger control on exchange than the ionic charge (smaller charge).

**Hibonite (Ca, Ce)(Al, Ti, Si, Mg)$_{12}$O$_{19}$**: The hibonite structure is complex and dominated by polyhedral layers perpendicular to the c-axis. $Ca^{+2}$ occupies a 12-coordinated polyhedron while $Al^{+3}$ is distributed over five M sites where M1 is a regular octahedron, M2 is a trigonal dipyramid, M3 is a tetrahedron, M4 is a trigonally distorted octahedron, and M5 is a strongly distorted octahedron (Li et al., 2016). Although Ti in general occupies M2 sites, depending on varying conditions of $fO_2$, $Ti^{+3}$ occupies the M1, M2, and M5 and $Ti^{+4}$ prefers to occupy M1, M2, M4, and M5 sites. Mg often occupies the M3 tetrahedral site. Previous crystallographic studies in meteoritic hibonites (Burns and Burns, 1984 and references therein) suggested that there is a charge balancing coupled substitution of $Mg^{+2} + Ti^{+4}$ for $2(Al^{+3} + V^{+3} + Cr^{+3})$. Significant Si contents of some meteoritic hibonites suggested that $Si^{+4}$ would be associated with the $Mg^{+2}$ site of the substitution and thereby permit additional Ti to exist as $Ti^{+3}$ ions along with $Ti^{+4}$ (Beckett and Stolper, 1994).

Our experiments show that most trace and REEs are incompatible in hibonite with varying degrees of incompatibility (Table 4). Our data is consistent with existing data in that hibonite is LREE enriched (Drake and Boynton, 1988; Kennedy et al, 1997, their Fig. 1a) (Fig. 11, Table 4). The low total number of experiments makes a detailed understanding of the temperature or compositional dependencies difficult. However, we can make some generalizations with regards to the behavior of groups of trace elements. The light elements Li, Be and B are all incompatible in hibonite, and do not significantly substitute for Al. The HFSE Ti, Zr, Hf, Nb, Ta and Th are moderately incompatible and most likely occupy the M sites. Sr and Ca are moderately incompatible (Ca partition coefficient is <1, even though it is a major component in the mineral), and Ba is highly incompatible in the twelve-fold site.

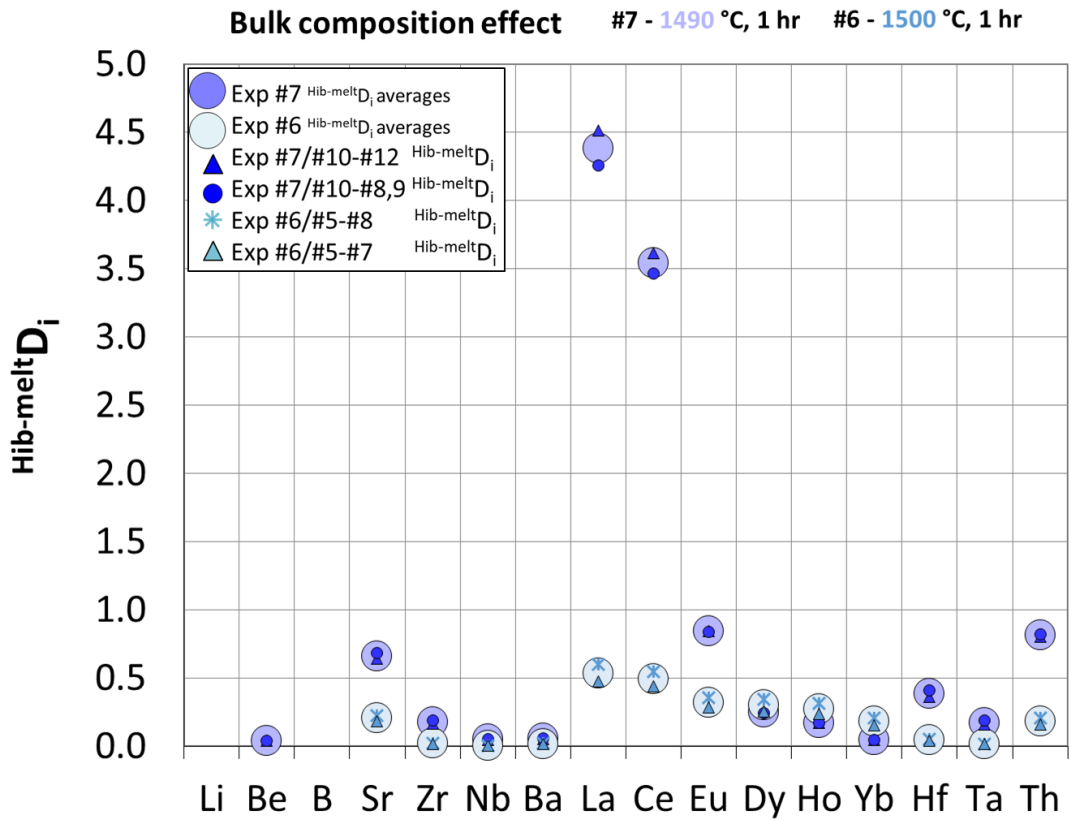

**Figure 11.** Effect of 2 different bulk compositions on trace and rare earth element partitioning between hibonite and melt ($^{Hib\text{-}melt}D_i$) for experiments #7 at 1490 °C, 1 hr; #6 at 1500 °C, 1 hr. Large filled and light colored circles show the average $^{Hib\text{-}melt}D_i$ for the given experiment. Small filled and dark colored symbols show the $^{Hib\text{-}melt}D_i$ from individual mineral-melt pair of the same colored experiment.

In general, the exchange for REE$^{+3}$ in the hibonite structure is with Al$^{+3}$, similar to grossite and gehlenite (Fig. 12). However, in contrast to grossite and gehlenite, hibonite has a more complex structure with five different sites for Al where three have octahedral coordination (M1, M4, and M5), one has tetrahedral coordination (M3), and one site ideally has trigonal bipyramidal coordination (M2) (Li et al., 2016). This structural complexity might be affected more by compositional changes in the melt rather than by temperature.

Current experiments show that $D_{La}$ increases from 0.6 to 4.4 between experiment #6 (average melt Al$_2$O$_3$: 42.90 wt%; T: 1500°C, P: 5kbar, $t$: 1hr, $f$O$_2$: IW+1) and # 7 (average melt

Al$_2$O$_3$: 48.64 wt%; T: 1490°C, P: 5kbar, *t*: 1hr, *f*O$_2$: IW+1) with a subtle 10°C difference in temperature at the same pressure, *f*O$_2$, and run duration. In contrast with Drake and Boynton (1998) and Kennedy et al. (1997) we are able to constrain the magnitude of change in bulk composition or average Al$_2$O$_3$ of melt on partitioning behavior of La between melt and hibonite. Obviously more experimental data will be necessary to understand individual effects of temperature, *f*O$_2$, and bulk composition.

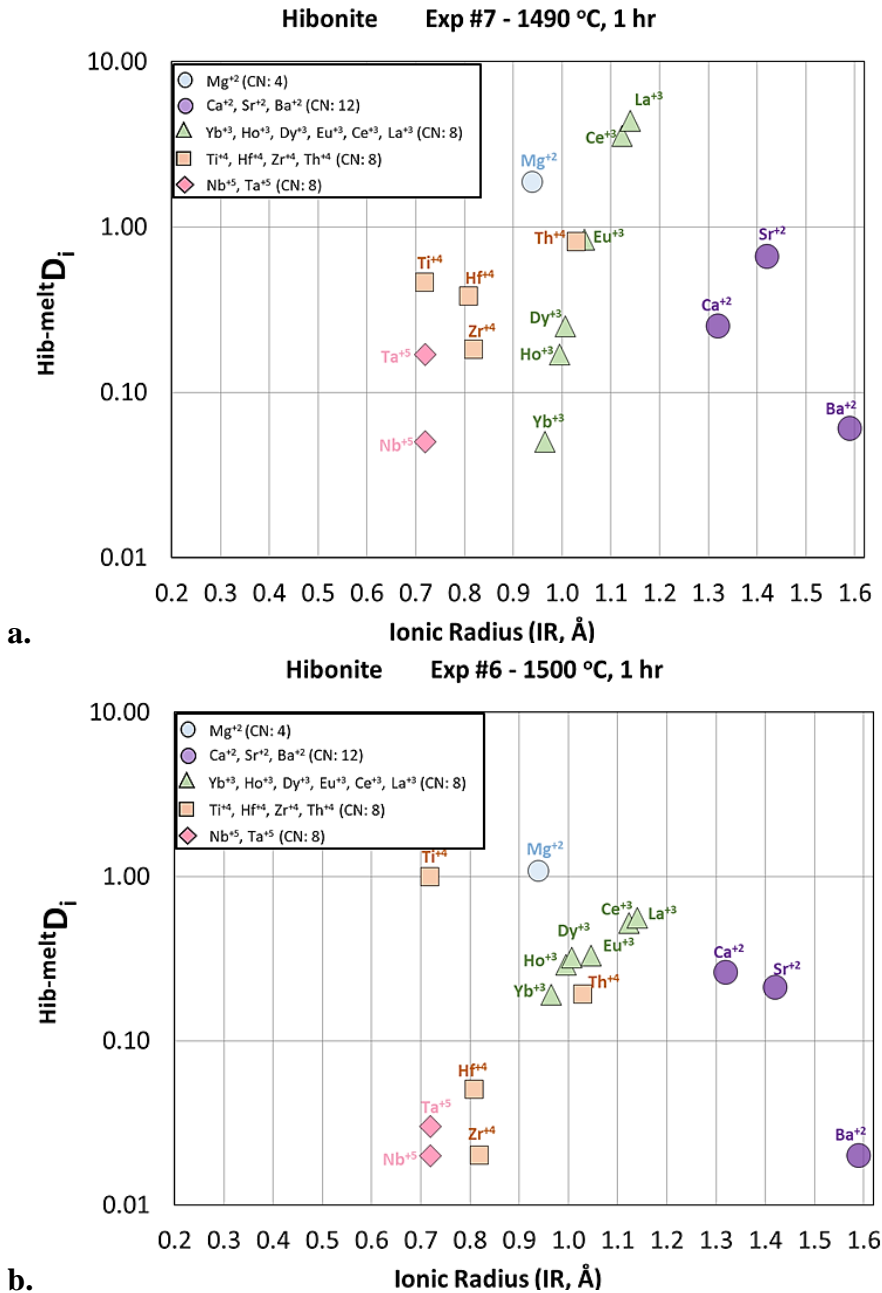

**Figure 12.** Onuma diagram for trace element partitioning between hibonite and melt for two different bulk compositions for $^{\text{Hib-melt}}D_i$ for experiments **a.** #7 at 1490 °C, 1 hr and **b.** #6 at 1500 °C, 1 hr. Ionic radii reflect values for CN = 4, CN = 8, and CN = 12 cations (Shannon, 1976).

Hibonite condenses at relatively high temperature (Yoneda and Grossman, 1995; Ebel, 2006), and has a structure which can incorporate significant amounts of polyvalent elements such as Ti and Fe. Therefore its substitutions might be sensitive to the fugacity conditions of the nebular gas (Lodders, 2003), and is critical to understanding the REE budget of the early nebula (Drake and Boynton, 1988; MacPherson and Davis, 1994).

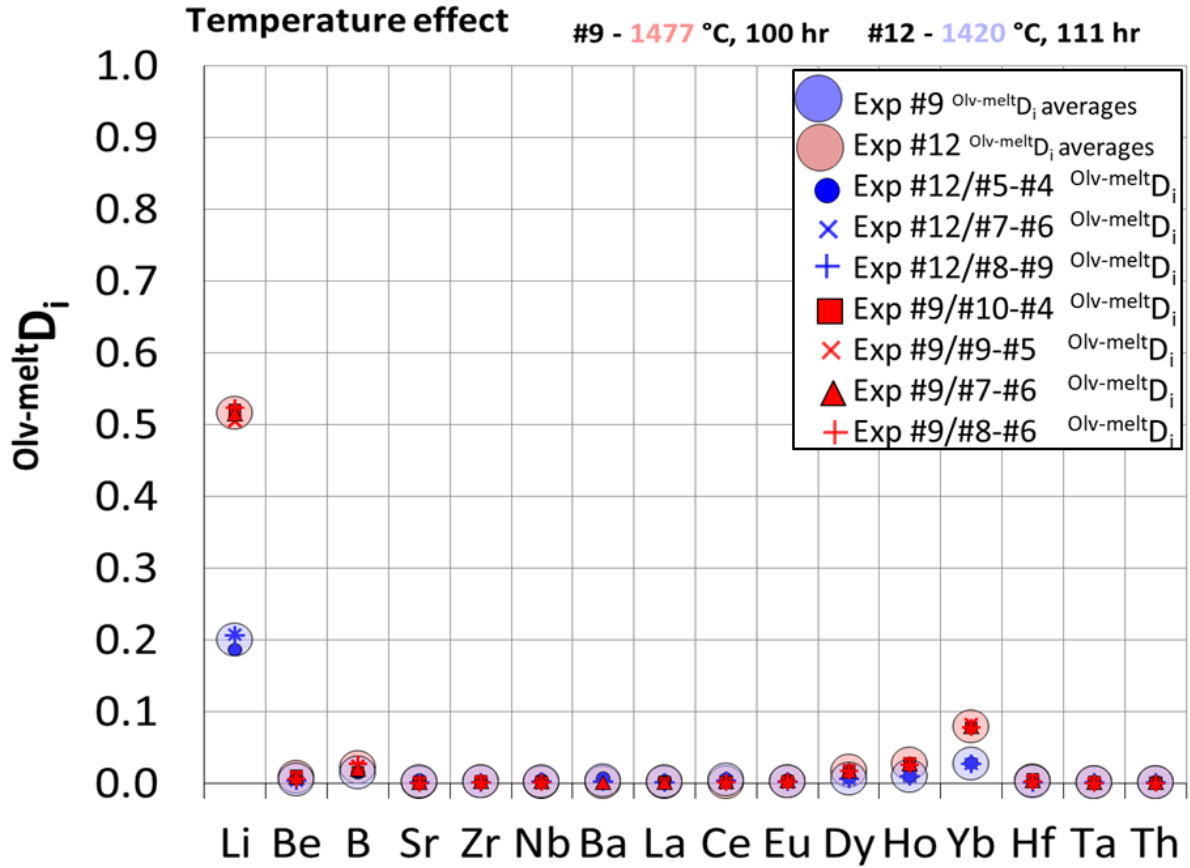

**Figure 13.** Temperature control on trace and rare earth element partitioning between olivine and melt ($^{Olv-melt}D_i$) using bulk composition pairs from experiments #9 at 1477 °C, 100 hrs and #12 at 1420 °C, 111 hrs. Large filled and light colored circles show the average $^{Olv-melt}D_i$ for the given experiment. Small filled and dark colored symbols show the $^{Olv-melt}D_i$ from individual mineral-melt pair of the same colored experiment. Red indicates high temperature runs while blue correlates with lower temperature.

**Olivine** $(Mg,Fe)_2SiO_4$: The structure of olivine consists of two octahedral sites (larger M2 and smaller M1, with +2 cations) and one tetrahedral site normally occupied by Si. When Ca is present it normally occupies the larger M2 site. The light elements Li, Be, and B are assumed to occupy the smaller M1 site. The larger REE and HSFE substitute into the M2 site, paired with cation vacancies (Nielsen et al., 1992; Bédard, 2005; Evans et al., 2008).

We report only the olivine data from experiments #12 and #9 (Table 5). In contrast to the higher temperature experiments (#12 and #9), in the lower temperature experiments (#15 and #21) the higher modal percentage of crystals made obtaining single phase analysis of olivine difficult. Because of the low partition coefficients for most studied elements in olivine, determination of partition coefficients is particularly sensitive to contamination with glass or other phases (Nielsen et al., 2017). Measured Ds from Be to Th in #15 and #21 average 0.021 and 0.022 respectively, with standard deviations 0.001. Such flat patterns are inconsistent with results from the higher temperature experiments, strongly suggesting contamination with glass. Such patterns are inconsistent with results from the higher T experiments, strongly suggesting contamination of olivine analyses by glass.

All trace elements measured in experiments #9 and #12 are incompatible in olivine (Fig. 13). This is consistent with the large body of experimental data that documents the details of olivine/melt partitioning (e.g., Evans et al., 2008; Mallman and O'Neill, 2013). The degree to which the light elements, REE and the HSFE are incompatible means that olivine controls the trace element budget of CAIs only indirectly (e.g., by dilution or exclusion). However, one important observation with respect to these experiments is that Yb is less incompatible than the surrounding REE (Fig. 14). Yb anomalies (e.g. Hiyagon et al., 2011) have been attributed to the greater volatility of Yb. However, our experiments were in sealed capsules, and volatility should not have been an issue. Another possibility may be the presence of $Yb^{+2}$, which may be sensitive to both temperature and $fO_2$. Ytterbium anomalies are present in many CAI inclusions and the $fO_2$ of such environments likely falls within the conditions of these experiments (Cartier et al., 2014). For our experiments, we did not observe a change in the partitioning behavior of tetravalent $Zr^{+4}$ in olivine depending on the CaO content of the melt as suggested earlier by Evans et al. (2008) (Table 5).

Among all studied trace and rare earth elements, $^{Olivine-Melt}D_i$ for Li is found to be the highest (0.52) in experiment #9 at 1477°C and IW+1. Previous studies have shown Li partitioning between olivine and silicate melt to be coupled with $Al^{+3}$ (Suzuki and Akaogi, 1995; Taura et al., 1998). The most favorable mechanism for substitution of a univalent cation in forsterite is suggested to involve pairing a +3 cation in the M2 site with a +1 cation in the M1 site (Purton et al., 1997). Their observations suggested the following mechanism for incorporating Li into the olivine structure:

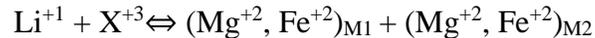

$$Li^{+1} + X^{+3} \Leftrightarrow (Mg^{+2}, Fe^{+2})_{M1} + (Mg^{+2}, Fe^{+2})_{M2}$$

The partitioning of Li between olivine and silicate melt was found to be affected by the FeO content in the olivine where at high temperatures $^{Olivine-Melt}D_{Li}$ increases with increasing FeO content in olivine (Brenan et al., 1998; Taura et al., 1998; Blundy and Dalton, 2000; Ottolini et al., 2009). Comparison of $^{Olivine-Melt}D_{Li}$ between the Caciagli et al. (2011) and Zanetti et al. (2004) experiments shows that regardless of the FeO content of olivine, $fO_2$ is a stronger control on partitioning. In spite of similar but relatively (compared to CAI) high FeO content of the olivine (19 wt.%), $^{Olivine-Melt}D_{Li}$ was much lower in reducing conditions (QFM-2) (Zanetti et al., 2004) compared to experiments buffered at QFM+4 (Caciagli et al., 2011). Such reducing conditions caused lower $Fe^{+3}$ contents and therefore less favorable conditions for $Li^{+1}$ substitution regardless of the high total FeO content of the olivine. This observation was later confirmed by the experiments conducted by Grant and Wood (2010) with a synthetic Fe-free olivine where the $^{Olivine-Melt}D_{Li}$ are well below other experimentally determined values.

## 5.2. Implications of the experimental results with respect to the evolution of CAIs

Our results suggest that, if clinopyroxene and perovskite are not present, the trace element budget of CAIs may be dominated by the three major Al-bearing phases including melilite (gehlenite), hibonite, and grossite. For all studied trace elements (LILE, REE and HFSE), gehlenite has higher partition coefficients than hibonite and to a lesser extent, grossite. If these three phases are present in a CAI, and clinopyroxene and perovskite are not, most of the trace elements would be hosted within them.

**Table 5.** Major (wt%) and trace element composition (ppm) of olivine and melt pairs from experiments #12 and #9 with average olivine-melt partition coefficients. Detection limits (ppm) for the measured trace elements are given on the right. Bdl: below detection limit. Numbers within the parentheses are errors (1σ).

| Exp #12 | Pair 1 | | Pair 2 | | Pair 3 | | | | | | |
|---|---|---|---|---|---|---|---|---|---|---|---|
| Phase | olivine (#5) | melt (#4) | olivine (#7) | melt (#6) | olivine (#8) | melt (#9) | | | | | |
| $SiO_2$ (wt%) | 40.07 (0.81) | 48.61 (0.65) | 39.74 (0.81) | 48.61 (0.65) | 40.07 (0.81) | 47.40 (0.65) | | | | | |
| $TiO_2$ | 0.04 (0.01) | 0.69 (0.06) | 0.04 (0.01) | 0.69 (0.06) | 0.04 (0.01) | 0.67 (0.06) | $^{Olv-Melt}D$ (averages from pairs) | | | | |
| $Al_2O_3$ | 0.19 (0.01) | 16.77 (0.66) | 0.17 (0.01) | 16.77 (0.66) | 0.19 (0.01) | 17.60 (0.66) | Li | 0.200 (0.012) | | | |
| FeO | 0.14 (0.02) | 0.17 (0.02) | 0.15 (0.02) | 0.17 (0.02) | 0.14 (0.02) | 0.15 (0.02) | Be | 0.005 (0.001) | | ~ Detection limits | |
| MgO | 57.47 (0.16) | 21.58 (1.34) | 57.44 (0.16) | 21.58 (1.34) | 57.47 (0.16) | 21.87 (1.34) | B | 0.015 | | olivine | melt |
| CaO | 0.36 (0.01) | 12.74 (0.62) | 0.36 (0.01) | 12.74 (0.62) | 0.36 (0.01) | 12.82 (0.62) | Mg | 2.651 (0.014) | Li (ppm) | 0.42 | 0.68 |
| $Cr_2O_3$ | 0.05 (0.01) | 0.06 (0.00) | 0.05 (0.01) | 0.06 (0.00) | 0.05 (0.01) | 0.06 (0.00) | Ti | 0.059 (0.011) | Be | 0.31 | 0.59 |
| Li (ppm) | 5.01 (0.43) | 26.84 (0.78) | 5.82 (0.62) | 28.06 (0.81) | 5.74 (0.53) | 27.74 (0.73) | Ca | 0.028 (0.010) | B | 1.11 | 1.67 |
| Be | 1.83 (0.41) | 274.73 (2.11) | 1.16 (0.651) | 259.32 (3.32) | 1.12 (0.37) | 264.55 (2.57) | Sr | 0.003 (0.001) | Sr | 0.08 | 0.13 |
| B | 1.27 (1.01) | 86.13 (2.07) | bdl | 86.39 (2.27) | bdl | 88.05 (1.91) | Zr | 0.002 (0.001) | Zr | 0.10 | 0.17 |
| Sr | 0.74 (0.07) | 155.64 (0.44) | 0.37 (0.10) | 155.72 (0.72) | 0.31 (0.10) | 157.75 (0.53) | Nb | 0.003 (0.001) | Nb | 0.02 | 0.03 |
| Zr | 1.05 (0.10) | 315.99 (0.86) | 0.56 (0.12) | 311.13 (1.41) | 0.55 (0.12) | 314.73 (1.05) | Ba | 0.005 (0.001) | Ba | 0.26 | 0.40 |
| Nb | 1.11 (0.03) | 198.52 (0.52) | 0.43 (0.05) | 195.11 (0.86) | 0.45 (0.04) | 196.98 (0.63) | La | 0.002 (0.001) | La | 0.02 | 0.03 |
| Ba | 1.52 (0.23) | 205.57 (0.92) | 0.68 (0.30) | 206.85 (1.43) | 0.70 (0.30) | 207.60 (1.08) | Ce | 0.005 (0.001) | Ce | 0.01 | 0.02 |
| La | 0.47 (0.02) | 113.89 (0.31) | 0.19 (0.03) | 112.18 (0.52) | 0.18 (0.03) | 113.64 (0.38) | Eu | 0.003 (0.001) | Eu | 0.11 | 0.01 |
| Ce | 0.75 (0.02) | 116.94 (0.31) | 0.54 (0.30) | 116.35 (0.53) | 0.51 (0.03) | 118.43 (0.39) | Dy | 0.007 (0.001) | Dy | 0.18 | 0.02 |
| Eu | 0.57 (0.02) | 116.81 (0.33) | 0.25 (0.03) | 116.96 (0.55) | 0.27 (0.03) | 117.18 (0.41) | Ho | 0.010 (0.001) | Ho | 0.23 | 0.01 |
| Dy | 0.83 (0.02) | 104.22 (0.35) | 0.69 (0.07) | 102.27 (0.59) | 0.64 (0.05) | 104.18 (0.43) | Yb | 0.028 (0.001) | Yb | 0.79 | 0.01 |
| Ho | 1.25 (0.02) | 116.78 (0.30) | 1.15 (0.06) | 115.37 (0.50) | 1.15 (0.05) | 118.16 (0.37) | Hf | 0.003 (0.001) | Hf | 0.10 | 0.03 |
| Yb | 3.24 (0.05) | 114.42 (0.39) | 3.17 (0.19) | 113.06 (0.65) | 3.22 (0.15) | 115.17 (0.48) | Ta | 0.002 (0.001) | Ta | 0.03 | 0.01 |
| Hf | 0.40 (0.02) | 120.52 (0.37) | 0.28 (0.04) | 118.51 (0.62) | 0.24 (0.03) | 120.93 (0.46) | Th | 0.002 (0.001) | Th | 0.01 | 0.01 |
| Ta | 0.54 (0.01) | 162.81 (0.41) | 0.24 (0.02) | 160.10 (0.68) | 0.24 (0.02) | 161.57 (0.50) | | | | | |
| Th | 0.29 (0.01) | 98.15 (0.27) | 0.13 (0.02) | 96.65 (0.45) | 0.13 (0.01) | 98.12 (0.33) | | | | | |

| Exp #9 | Pair 1 | | Pair 2 | | Pair 3 | | Pair 4 | | | | | |
|---|---|---|---|---|---|---|---|---|---|---|---|---|
| Phase | olivine (#10) | melt (#4) | olivine (#9) | melt (#5) | olivine (#8) | melt (#6) | olivine (#7) | melt (#6) | | | | |
| $SiO_2$ (wt%) | 40.99 (0.77) | 47.54 (0.31) | 42.39 (0.77) | 47.54 (0.31) | 41.13 (0.77) | 47.10 (0.31) | 42.39 (0.77) | 47.10 (0.31) | | | | |
| $TiO_2$ | 0.02 (0.01) | 0.64 (0.05) | 0.02 (0.01) | 0.64 (0.05) | 0.02 (0.01) | 0.67 (0.05) | 0.02 (0.01) | 0.67 (0.05) | $^{Olv-Melt}D$ (averages from pairs) | | | |
| $Al_2O_3$ | 0.12 (0.04) | 14.51 (1.00) | 0.16 (0.04) | 14.51 (1.00) | 0.09 (0.04) | 15.92 (1.00) | 0.16 (0.04) | 15.92 (1.00) | Li | 0.516 (0.009) | | | |
| FeO | 0.25 (0.01) | 0.35 (0.01) | 0.25 (0.01) | 0.35 (0.01) | 0.26 (0.01) | 0.34 (0.01) | 0.25 (0.01) | 0.34 (0.01) | Be | 0.009 (0.002) | | ~ Detection limits | |
| MgO | 57.54 (0.19) | 25.50 (0.16) | 57.82 (0.19) | 25.50 (0.16) | 57.90 (0.19) | 25.28 (0.16) | 57.82 (0.19) | 25.28 (0.16) | B | 0.022 (0.005) | | olivine | melt |
| CaO | 0.31 (0.01) | 11.57 (0.03) | 0.31 (0.01) | 11.57 (0.03) | 0.31 (0.01) | 11.61 (0.03) | 0.31 (0.01) | 11.61 (0.03) | Mg | 2.275 (0.010) | Li (ppm) | 0.51 | 1.90 |
| $Cr_2O_3$ | 0.06 (0.01) | 0.10 (0.00) | 0.06 (0.01) | 0.10 (0.00) | 0.06 (0.01) | 0.10 (0.00) | 0.06 (0.01) | 0.10 (0.00) | Ti | 0.031 (0.009) | Be | 1.69 | 0.99 |
| Li (ppm) | 37.71 (5.85) | 72.31 (1.79) | 37.60 (3.50) | 74.49 (2.13) | 39.31 (4.93) | 75.13 (1.98) | 38.71 (6.72) | 75.13 (1.98) | Ca | 0.027 (0.010) | B | 1.46 | 1.93 |
| Be | 2.87 (1.35) | 245.72 (2.18) | 2.15 (1.75) | 251.34 (2.17) | 1.54 (1.64) | 255.47 (2.40) | 2.43 (2.12) | 255.47 (2.40) | Sr | 0.002 (0.001) | Sr | 0.11 | 0.18 |
| B | 1.33 (0.90) | 78.66 (2.20) | 1.98 (1.35) | 79.90 (1.84) | 2.26 (1.60) | 81.55 (1.45) | 1.68 (1.21) | 81.55 (1.45) | Zr | 0.003 (0.001) | Zr | 0.10 | 0.16 |
| Sr | 0.26 (0.16) | 115.68 (0.38) | 0.14 (0.07) | 114.57 (0.37) | 0.11 (0.06) | 116.25 (0.39) | 0.25 (0.15) | 116.25 (0.39) | Nb | 0.002 (0.001) | Nb | 0.08 | 0.03 |
| Zr | 0.61 (0.29) | 149.30 (0.48) | 0.38 (0.29) | 151.15 (0.50) | 0.33 (0.29) | 152.75 (0.51) | 0.43 (0.30) | 152.75 (0.51) | Ba | 0.002 | Ba | 0.10 | 1.16 |
| Nb | 0.41 (0.09) | 145.39 (0.40) | 0.06 (0.02) | 144.16 (0.40) | 0.15 (0.06) | 145.45 (0.41) | 0.42 (0.07) | 145.45 (0.41) | La | 0.002 | La | 0.11 | 0.06 |
| Ba | bdl | 126.05 (1.36) | bdl | 125.84 (1.33) | bdl | 130.74 (1.35) | 0.16 (0.12) | 130.74 (1.35) | Ce | 0.002 (0.001) | Ce | 0.06 | 0.03 |
| La | 0.19 (0.13) | 82.71 (0.25) | bdl | 83.68 (0.26) | bdl | 84.57 (0.26) | 0.14 (0.11) | 84.57 (0.26) | Eu | 0.003 (0.001) | Eu | 0.14 | 0.07 |
| Ce | 0.25 (0.07) | 85.69 (0.25) | 0.07 (0.02) | 86.48 (0.25) | 0.13 (0.04) | 87.29 (0.26) | 0.28 (0.06) | 87.29 (0.26) | Dy | 0.018 (0.001) | Dy | 0.22 | 0.03 |
| Eu | 0.35 (0.01) | 81.09 (0.26) | 0.18 (0.03) | 81.90 (0.26) | 0.33 (0.13) | 83.32 (0.27) | 0.33 (0.13) | 83.32 (0.27) | Ho | 0.028 (0.001) | Ho | 0.26 | 0.01 |
| Dy | 1.52 (0.24) | 82.62 (0.32) | 1.58 (0.16) | 84.15 (0.33) | 1.52 (0.22) | 86.59 (0.34) | 1.51 (0.29) | 86.59 (0.34) | Yb | 0.080 (0.003) | Yb | 0.86 | 0.03 |
| Ho | 2.26 (0.28) | 78.82 (0.22) | 2.30 (0.10) | 81.05 (0.23) | 2.16 (0.30) | 83.72 (0.24) | 2.31 (0.38) | 83.72 (0.24) | Hf | 0.004 (0.001) | Hf | 0.10 | 0.03 |
| Yb | 7.43 (1.11) | 94.59 (0.37) | 8.04 (0.34) | 96.86 (0.38) | 7.64 (0.79) | 98.27 (0.39) | 7.82 (1.26) | 98.27 (0.39) | Ta | 0.002 (0.001) | Ta | 0.03 | 0.01 |
| Hf | 0.56 (0.12) | 86.84 (0.31) | 0.35 (0.05) | 90.24 (0.32) | 0.31 (0.12) | 92.24 (0.33) | 0.37 (0.10) | 92.24 (0.33) | Th | 0.002 (0.001) | Th | 0.02 | 0.01 |
| Ta | 0.31 (0.04) | 103.37 (0.28) | 0.05 (0.01) | 105.36 (0.29) | 0.06 (0.01) | 106.99 (0.30) | 0.17 (0.03) | 106.99 (0.30) | | | | | |
| Th | 0.14 (0.03) | 64.16 (0.20) | bdl | 66.21 (0.20) | 0.04 (0.02) | 68.09 (0.21) | 0.09 (0.02) | 68.09 (0.21) | | | | | |

Experiments #6 and #7 are multiply saturated with grossite, gehlenite, and hibonite. This provides us with the opportunity to calculate partition coefficients for solid-solid equilibria. Calculated $^{Grossite-Hibonite}D_i$, $^{Grossite-Gehlenite}D_i$, and $^{Gehlenite-Hibonite}D_i$ from experiment #7 (Fig. 14a) demonstrate that when all three phases are present, LILE (Sr, Ba); REEs (La, Ce, Eu, Dy, Ho, Yb); and HFSE (Th) prefer to be concentrated in gehlenite over hibonite and grossite. This is even more so for HREEs (Dy, Ho, and Yb) with $^{Gehlenite-Hibonite}D_i$ values ranging between 12-20. For the same composition (Exp #7) Zr, Nb, and Hf prefer grossite over gehlenite and hibonite. Experiment #6, which is multiply saturated with gehlenite and hibonite shows the same pattern as experiment #7, where Sr, Zr, Nb, Ba, La-Yb, Hf, Ta, Th partition into gehlenite over hibonite. The different $Al_2O_3$ content of the melt in #7 and #6 represents another compositional control over solid-solid state equilibria. $Al_2O_3$ abundance in experiment #7 (average melt $Al_2O_3$: 48.64 wt%) is significantly higher than in #6 (average melt $Al_2O_3$: 42.90 wt%) at approximately the same temperature (difference of 10°C) and the same pressure (5kbar). The $fO_2$ (IW+1), and run duration (1hr) correlate with increases in the partitioning of Nb, HREEs (Dy, Ho, Yb) and Th into gehlenite compared to hibonite, while low $Al_2O_3$ of the melt correlates with lower $^{Gehlenite-Hibonite}D_i$ for Sr, Zr, Ba, and LREEs (La, Ce, Eu).

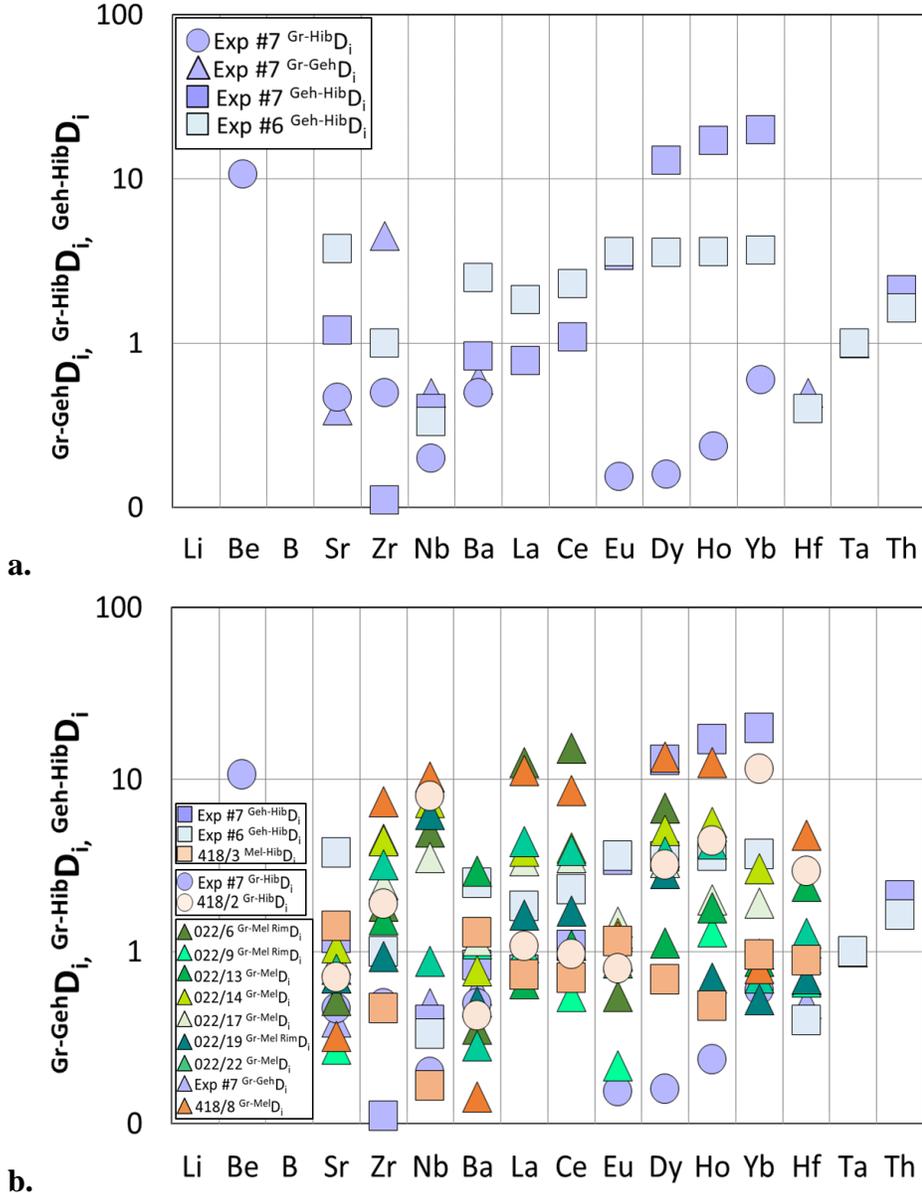

**Figure 14 a.** Comparison of grossite-gehlenite ($^{Gr-Geh}D_i$), grossite-hibonite ($^{Gr-Hib}D_i$), gehlenite-hibonite ($^{Geh-Hib}D_i$) partition coefficients from experiments #7 at 1500 °C, 1 hr and #6 at 1490 °C, 1 hr. b. Comparison between experimentally determined equilibrium partition coefficients for trace and rare earth elements from #6 and #7 with measured partition coefficients of CAIs from CR chondrites Acfer 182 and Acfer 059-El Djouf 001 (Weber et al., 1995).

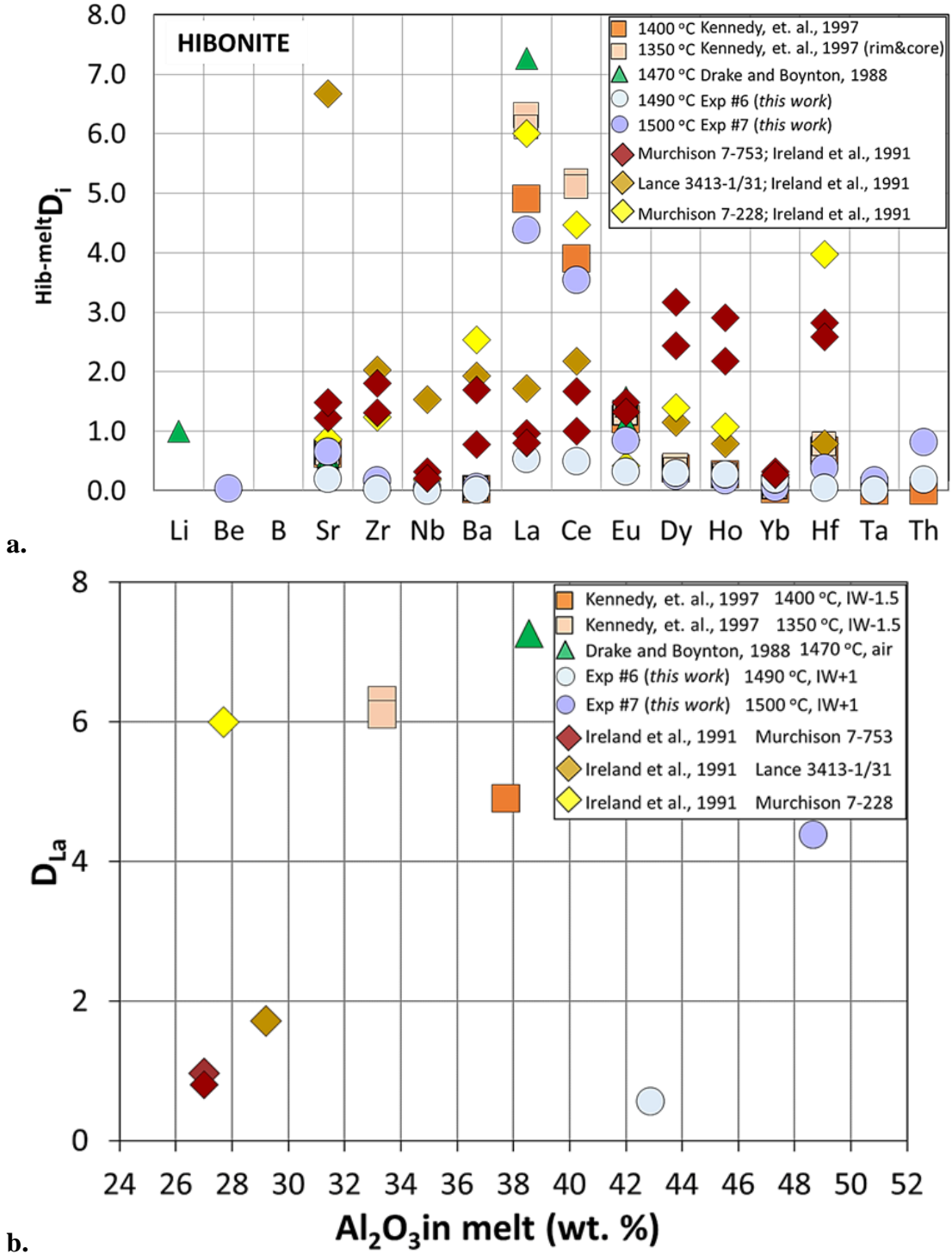

**Figure 15 a.** Comparison of hibonite-melt partition coefficients ($^{\text{Hib-melt}}D_i$) from Drake and Boynton (1998); Kennedy et al. (1997); and Ireland et al. (1991) with the equilibrium partitioning experiments from this work (Exp #7 and Exp#6). **b.** $^{\text{Hib-melt}}D_{\text{La}}$ versus $Al_2O_3$ content (wt%) in melts from Drake and Boynton (1998); Kennedy et al. (1997); Ireland et al. (1991) and experiments #7 and #6 from this work.

We also compared the experimentally determined equilibrium mineral-mineral partition coefficients (Exp. #6, #7) with measured $^{\text{Melilite-Hibonite}}D_i$, $^{\text{Grossite-Hibonite}}D_i$, and $^{\text{Grossite-Melilite}}D_i$ of grossite-rich (up to 30 volume%) CAIs from CR chondrites Acfer 182 and Acfer 059-El Djouf 001 (Fig. 14b; Weber and Bischoff, 1994; Weber et al., 1995). They measured a large range of these $D_i$ for the 10 multiply saturated CAIs (their samples 022/6, 022/9, 022/13, 022/14, 022/17, 022/19, 022/22 and 418/2, 418/3, 418/8). The range of partitioning values exhibited by the natural samples falls mostly within the range of experimental values with significant differences (Figure 14b). Measured $^{\text{Grossite-Melilite}}D_i$ from these inclusions are significantly higher compared to experiments #6 and #7. Measured $^{\text{Melilite-Hibonite}}D_i$ for Sr, Zr, Ba, and Hf are higher than Al-rich melt (Exp #7) but lower than Exp #6. Refractory lithophiles and significant number of REEs (Nb, La, Ce, Eu, Dy, Ho, and Yb) have lower $^{\text{Melilite-Hibonite}}D_i$ than in both experiments and therefore likely deviate significantly from equilibrium conditions. $^{\text{Grossite-Hibonite}}D_i$ exist only for one CAI (418/2) and is higher than experiment #7 for all studied trace elements.

We compared experimentally determined hibonite-melt partition coefficients to measured $^{\text{Hibonite-Melt}}D_i$, two from CM2 Murchison (7-228) and one from CO3 Lancé (3413-1/31) from Ireland et al. (1991), and with experimentally determined $^{\text{Hibonite-Melt}}D_i$ (Kennedy et al., 1997; Drake and Boynton, 1988) (Fig. 15). Both Murchison and Lancé have euhedral hibonite-bearing spherules and silicate glass. When compared to Murchison and Lancé, all studied trace elements (with the exception of La and Ce) from experiments #6 and #7 have lower hibonite-melt partitioning. The Kennedy et al. (1997) and Drake and Boynton (1998) experiments have higher $^{\text{Hibonite-Melt}}D_i$ for La and Ce than our experiments #6 and #7 (Fig. 15a), but lower $^{\text{Hibonite-Melt}}D_i$ for Eu, Yb, and Sr. Our experiments are higher in $Al_2O_3$ compared to both Murchison and Lancé CAI glass and to all the Kennedy et al., (1997) and Drake and Boynton (1988) experiments. Kennedy et al. (1997) conducted experiments at reduced conditions compared to those reported here, while the experiments of Drake and Boynton (1998) were conducted in air. Without data with either one or the other of the variables held constant, it is not possible to constrain the temperature, composition, and $fO_2$ controls on partitioning. Such experiments are difficult to conduct due to the relatively narrow ranges of stability of Al-rich CAI phases.

Although our experiments provide new temperature and bulk composition constraints on partitioning behavior of investigated trace elements, the application of this new information is limited by the small crystal size characteristic of most CAIs and the paucity of available single phase (mineral-melt pair) microanalytical data in the existing literature. Obtaining such information has been difficult because the majority of the trace and rare earth element analyses of the phases in CAIs as well as chondrules were performed with laser ICP-MS whose analytical volume exceeds the volume of the critical phases (e.g., for olivine-melt, Crapster-Pregont et al., 2018, on CVs; cf., Alexander, 1994, on OCs). Therefore, many published analyses may represent analytical volumes that include more than one phase.

We now have a wide range of trace element partition coefficients that cover a significant proportion of the compositional range of naturally occurring CAIs. The next step is to obtain new trace element data on CAI mineralogy at a higher degree of spatial resolution. This is required if we are to apply quantitative models using these new coefficients to establish conditions and the degree to which natural CAIs and/or chondrules obtained equilibrium.

# 6. CONCLUSIONS

New partition coefficients for melilite (gehlenite), hibonite, grossite, and olivine with melt have documented some of the compositional and temperature dependencies which should be applicable to a range of CAI-type compositions. We conceived a strategy to track variation of bulk composition in a compositional space relevant to CAI evolution and therefore expand on the limited existing database for partitioning for several elements and track this compositional variation through a range of temperatures relevant to CAI evolution. This new data also provides additional details (composition, temperature, and $fO_2$) that expand the existing experimental constraints showing that the trace element budget of CAIs will be dominated by the modal abundance of melilite, hibonite, grossite and to a lesser extent olivine and pyroxene (Cartier et al., 2014). For most trace element groups, the modal proportion of olivine was found to be a diluent because of the low olivine-melt partition coefficients. In order to move this research forward, new data must be collected on natural CAIs with better spatial resolution, where one can be confident that the analytical volume represents the composition of a single phase. This need is caused by the fine intergrowth and melt inclusion textures characteristic of most CAIs and related refractory inclusions.

**ACKNOWLEDGEMENTS**

The authors thank J. Beckett and Q-Z. Yin for detailed and thoughtful reviews that greatly helped to improve the quality and clarity of the manuscript. The authors are also grateful to A. Krot for the editorial handling of this manuscript. This work was supported by U.S. N.A.S.A. grant NNX10AI42G (DSE), and an AMNH Katherine Davis post-doctoral fellowship (GU). This research has made use of NASA's Astrophysics Data System Bibliographic Services.